\documentclass[aps,prd,preprint,nofootinbib]{revtex4-1}
\usepackage{amsmath,amssymb,graphics,graphicx,color}
\usepackage{scrextend}
\usepackage{hyperref}
\usepackage{float}
\usepackage{xcolor}
\usepackage[utf8]{inputenc}

\usepackage{multirow}
\usepackage{bigcenter}
\usepackage{caption}
\usepackage{subcaption}

\newcommand{\bib}[1]{Ref.~\cite{#1}}
\newcommand{\refs}[1]{Refs.~\cite{#1}}
\newcommand{\bibs}[1]{\cite{#1}}
\newcommand{\fig}[1]{Fig.~\ref{#1}}

\begin{document}

\title{Leptoquark search at the Forward Physics Facility}
\author{Kingman Cheung$^{a,b,c}$, Thong T.Q. Nguyen$^{d}$, C.J. Ouseph$^{a,b}$}
\affiliation{
  $^a$ Department of Physics, National Tsing Hua University, Hsinchu 30013,
  Taiwan\\
  $^b$ Center for Theory and Computation, National Tsing Hua University, Hsinchu 30013, Taiwan \\
  $^c$ Division of Quantum Phases and Devices, School of Physics,
  Konkuk University, Seoul 143-701, Republic of Korea\\
  $^d$ Insitute of Physics, Academia Sinica, Nangang, Taipei 11529, Taiwan
}

\date{\today}
\begin{abstract}
In this study, we calculate the sensitivity reach on the 
vector leptoquark (LQ) $U_1$ at the experiments proposed in 
Forward Physics Facility (FPF), including FASER$\nu$, FASER$\nu2$, 
FLArE (10 tons), and FLArE (100 tons) using the neutrino-nucleon
scattering ($\nu N \rightarrow \nu N'$ and $\nu N \rightarrow l N'$).
We cover a wide mass range of $10^{-3}$ GeV $\leq M_{LQ}\leq 10^4$ GeV.
The new result shows that the FLArE (100 tons) offers the best 
sensitivity to the LQ model.
The sensitivity curves for all the experiments follow a similar pattern 
with weakened sensitivities with the increment of the LQ mass. 
We combine the sensitivities obtained from the neutral- and 
charged-current interactions of the neutrinos.
\end{abstract}
\maketitle

\section{Introduction}
Hunting for new physics beyond the standard model (SM) 
is one of the major goals at various experiments in high-energy and intensity frontiers.
While the high-energy frontier has not found anything other than the discovery of the Higgs boson, 
the precision frontier, on the other hand, seemed to show some surprising results
on a number of B meson decays, 
and the muon anomalous magnetic dipole moment, although 
more data are needed to confirm.

The discrepancies existed between the SM predictions and the
experimental results for the flavor-changing neutral current rare 
decays of B mesons in $b \to s \ell^+ \ell^-$, in particular, the 
lepton-flavor universality violation in $B \to K$
transition observed by LHCb, expressed in terms of $R_K$ and $R_{K^{*}}$:
\begin{equation}
  R_K = \frac{ {\rm BR}( B \to K \mu^+ \mu^-) }{ {\rm BR}( B \to K e^+ e^-) } \;, \,\;\;\;
  R_{K*} = \frac{ {\rm BR}( B \to K^* \mu^+ \mu^-) }{ {\rm BR}( B \to K^* e^+ e^-) }
  \;.
\end{equation}
\footnote{
The discrepancies between the SM predictions and the
experimental results for the flavor-changing neutral current rare 
decays of B mesons in $b \to s \ell^+ \ell^-$ were as large as $3\sigma$
with the measurements \cite{LHCb:2021trn,LHCb:2017avl}
\begin{equation}
  R_K = 0.846\,^{+0.042}_{-0.039}\, ^{+0.013}_{-0.012} \;,  \;\;\;
  \mbox{for 1.1 GeV$^2  < q^2 < 6$ GeV$^2$} \;,
\end{equation}
\begin{equation}
  R_{K^*} = \left \{ \begin{array}{lr}
  0.66 \, ^{+0.11}_{-0.07} \, \pm 0.03 & 0.045 \;{\rm GeV}^2 < q^2 < 1.1 \; {\rm GeV}^2\,,\\
  0.69 \, ^{+0.11}_{-0.07} \, \pm 0.05 &  1.1 \;{\rm GeV}^2 < q^2 < 6.0\; {\rm GeV}^2\,,
  \end{array}
   \right .
\end{equation}
}
Nevertheless, the anomalies faded away from in
the most recent measurements \cite{LHCb:2022zom}, which showed
consistency with the SM predictions:
\begin{equation}
  R_{K} = \left \{ \begin{array}{lr}
  0.994 \, ^{+0.090}_{-0.082}\mathrm{(stat)}^{+0.029}_{-0.027}\mathrm{(syst)}  & \;{\rm low-}q^{2},\\
  0.949 \, ^{+0.042}_{-0.041}\mathrm{(stat)}^{+0.036}_{-0.035}\mathrm{(syst)} &\;{\rm central-}q^{2},
  \end{array}
   \right .
\end{equation}
\begin{equation}
  R_{K^{*}} = \left \{ \begin{array}{lr}
  0.927 \, ^{+0.042}_{-0.041}\mathrm{(stat)}^{+0.022}_{-0.022}\mathrm{(syst)}  & \;{\rm low-}q^{2},\\
  1.027 \, ^{+0.072}_{-0.068}\mathrm{(stat)}^{+0.027}_{-0.026}\mathrm{(syst)} &\;{\rm central-}q^{2}.
  \end{array}
   \right .
\end{equation}

Another set of observables related to the short-distance process $b \to c \ell\nu$ 
are \cite{HFLAV:2019otj}
\begin{eqnarray}
R_{D} &=& \frac{ {\rm BR}(B \to D \tau \nu) } { {\rm BR}(B \to D \ell \nu) } =
0.340 \pm 0.027 \pm 0.013\,,  \nonumber \\
R_{D^*} &=& \frac{ {\rm BR}(B \to D^* \tau \nu) } { {\rm BR}(B \to D^* \ell \nu) } =
0.295 \pm 0.011 \pm 0.008 \;, 
\end{eqnarray}
which still showed discrepancies from the SM. 

Another long-standing experimental anomaly is the muon anomalous moment
(aka. $g-2$).  The most recent muon $g-2$ measurement was performed by the 
E989 experiment at Fermilab, which  reported the new result \cite{Muong-2:2021ojo}
\begin{equation}
  \label{amu}
\Delta a_\mu = (25.1 \pm 5.9) \times 10^{-10} \;,
\end{equation}
which deviates at the level of $4.2\sigma$ from the known SM predictions 
before the recent lattice results. The recent lattice results 
\cite{Borsanyi:2020mff,Blum:2019ugy} made substantial
improvements in hadronic contributions to $g-2$ 
such that the deviation of the experimental
result only stands at about $1-2 \sigma$ level.

Leptoquark (LQ) models were suggested to explain some or all of the
above anomalies. Especially, it was shown in 
Refs.~\cite{Calibbi:2015kma,Calibbi:2017qbu,Angelescu:2021lln} that 
the isosinglet vector LQ $U_1$ can explain both 
$R_{K,K^*}$ and $R_{D,D^*}$, and in Ref.~\cite{Cheung:2022zsb} that
the isodoublet vector LQ $V_2$ provides a viable solution to
$R_{K,K^*}$, $R_{D,D^*}$, and muon $g-2$. On the other hand, other
LQ models can only explain one or some of the anomalies, unless
with more than 1 leptoquarks. 
In this work, we do not concern ourselves with the second and third-generation quark couplings, nor do we assume any flavor symmetries. We only consider the first-generation quark couplings as they are more relevant in neutrino-nucleon scattering. Our results will not reflect any constraints on the second or third-generation couplings.

Such LQs have been searched at the LHC with strong limits on the 
LQ mass via leptoquark pair production. The mass limits depend
on the decay channels of the LQs.  Nevertheless, such decays 
often make use of the decay into a quark and a charged lepton.
When the LQ decays into a quark plus a neutrino, the limits
are much weakened.

In this work, we investigate the effects of LQs via
neutrino-nucleon scattering at the FASER$\nu$, and the future 
FASER2$\nu$ and FLArE experiments. The ultimate plan is to have 
a Forward Physics Facility (FPF) \cite{Feng:2022inv}, which can
house a number of such experiments.
FASER$\nu$ is indeed running and taking data 
\cite{talk-2022,talk-2022(2)}.
Such experiments use
the energetic neutrinos produced by the decays of mesons (e.g. pions,
kaons) from the interaction point of the ATLAS experiment.
The most distinct feature is that the energy range can be as 
high as TeV, which provides an unprecedented energy scale of 
studying neutrino-nucleon and neutrino-electron scattering.

We focus on the isosinglet vector LQ $U_1$, though the results 
can be easily adapted to the isodoublet vector LQ $V_2$. 
An interesting feature of the LQ is that it can enhance 
both the charged- and neutral-current interactions, so it can
give better sensitivities than the $Z'$ interactions, which only 
enhance the neutral-current interactions \cite{Cheung:2021tmx}.
We calculate both the charged- and neutral-current scattering via the
LQ and obtain the sensitivities that one can obtain at FASER$\nu$,
FASER2$\nu$, and FLArE. 

The organization is as follows. In the next section, we describe
the interactions of the LQ that are relevant to our study. 
In Sec. III, we present the numerical results and the sensitivity
reach at various FPF experiments. 

Note that we are exploring very light leptoquarks with small
couplings to quarks and leptons. 
Nevertheless, the relevant couplings in our study 
are those of LQ with the first-generation quarks ($u$ and $d$) in 
the $\nu N$ scattering, while those explaining the $B$ anomalies 
concern mostly the third and second generations. Thus, we do not 
restrict the LQ couplings to those obtained in $B$ anomalies.

\section{Theoretical Setup}

\subsection{Leptoquark model}
Leptoquarks are predicted in many grand unified theories, which
couple to both quarks and leptons.
It was shown in \bib{Angelescu:2021lln} that 
the singlet vector leptoquark $U(1)$ with the SM quantum numbers 
({\bf 3}, {\bf 1}, 2/3) can explain both $R_{K,K^{*}}$ and 
$R_{D,D^{*}}$ anomalies.
The Yukawa interactions for $U_1$ can be written as 
\cite{DiLuzio:2018zxy, Baker:2019sli, Cornella:2021sby}:
\begin{equation}
\begin{split}
\mathcal{L}_{U_{1}}=&
\frac{g_{U}}{\sqrt{2} } [U^{\mu}_{1}\,
(\beta^{ij}_{L}\bar{q}^{i}_{L}\gamma_{\mu}l^{j}_{L}+
\beta^{ij}_R \bar{d}^{i}_{R}\gamma_{\mu}l^{j}_{R})+\mathrm{h.c.}],
\end{split}
\end{equation}
where $q_L, l_L, d_R, l_R $ denote the quark doublet, lepton doublet,
down-type quark singlet, and lepton singlet, respectively.
%
Here $i$ and $j$ denote the generation indices and 
$\beta^{ij}_{L/R}$ allow for generation mixing, and 
$g_U$ is the overall coupling strength.

We used the {\bf Vector Leptoquark models} from Refs. \cite{DiLuzio:2018zxy, Baker:2019sli, Cornella:2021sby}, which can be accessed from the FeynRules model database (\cite{Alloul:2013bka}). 
We choose $\beta^{2j}$ and $\beta^{3j}$ (for $j=1,2,3$) 
to be zero, such that the flavor-changing neutral currents 
among the first, the second, and the third-generation quarks 
are avoided. Also, only the first-generation-quark couplings
$\beta^{1j}$ are relevant to our study. 
Such choices may give rise to lepton-flavor changing processes, e.g., $\tau \to \mu$ decays, but, however, we expect them to be negligible,
as no second- or third-generation quarks are involved.

Note that although we use the leptoquark $U_1$ in our study, 
the cases with other leptoquarks would give results with 
similar orders of magnitude on the sensitivity reach. Unlike scalar leptoquarks, vector leptoquarks must have 
some extra degrees of freedom due to the UV sensitivity of the 
loop \cite{Cornella:2019hct}, where an example of UV completion was given.

\subsection{Neutrino-nucleon scattering via leptoquarks}

\begin{figure}[th!]
\centering
\begin{subfigure}{0.49\textwidth}
\centering
\includegraphics[width=\textwidth]{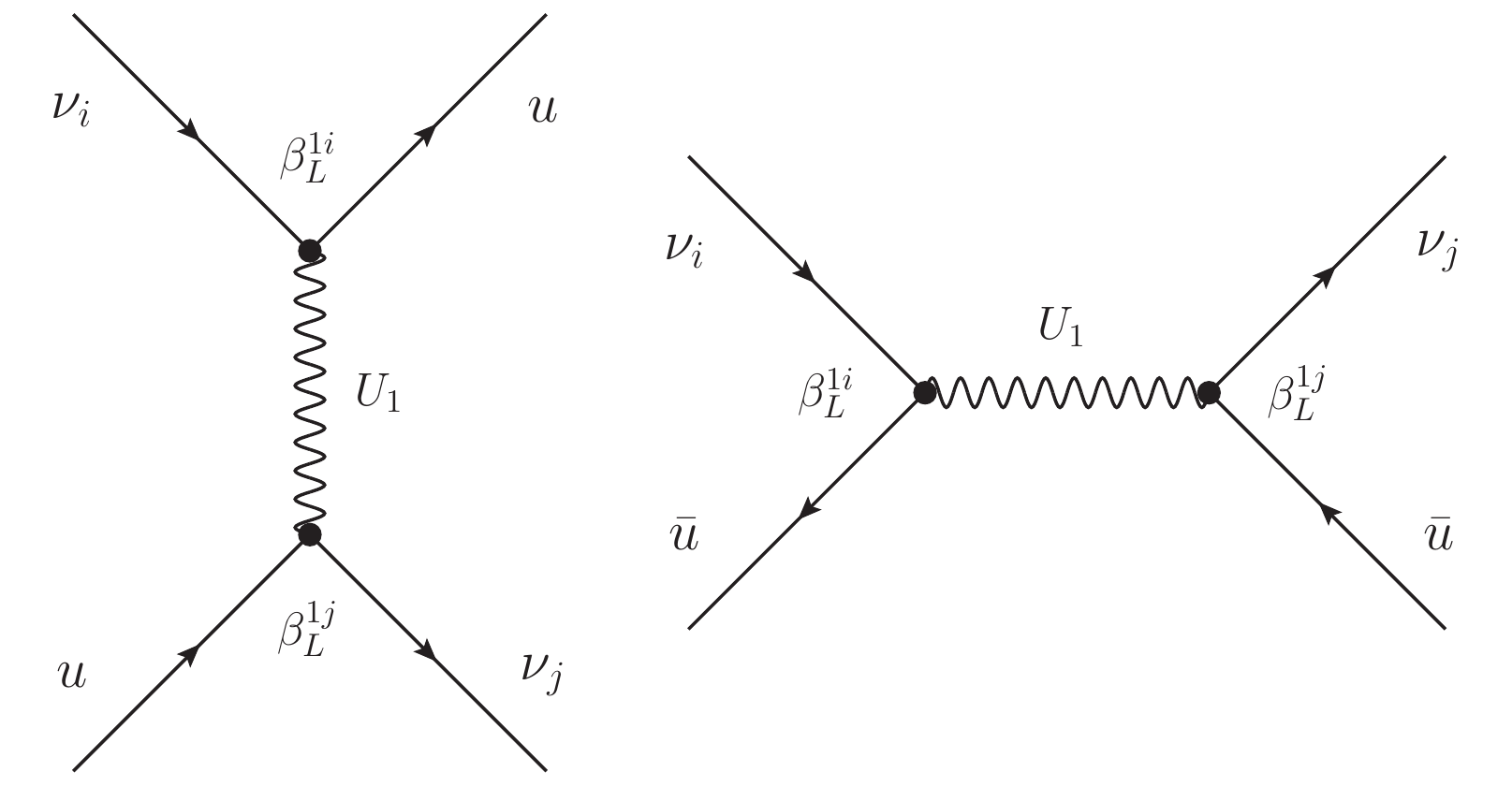}
\caption{Neutral Current}
\end{subfigure}
\hfill
\begin{subfigure}{0.49\textwidth}
\centering
\includegraphics[width=\textwidth]{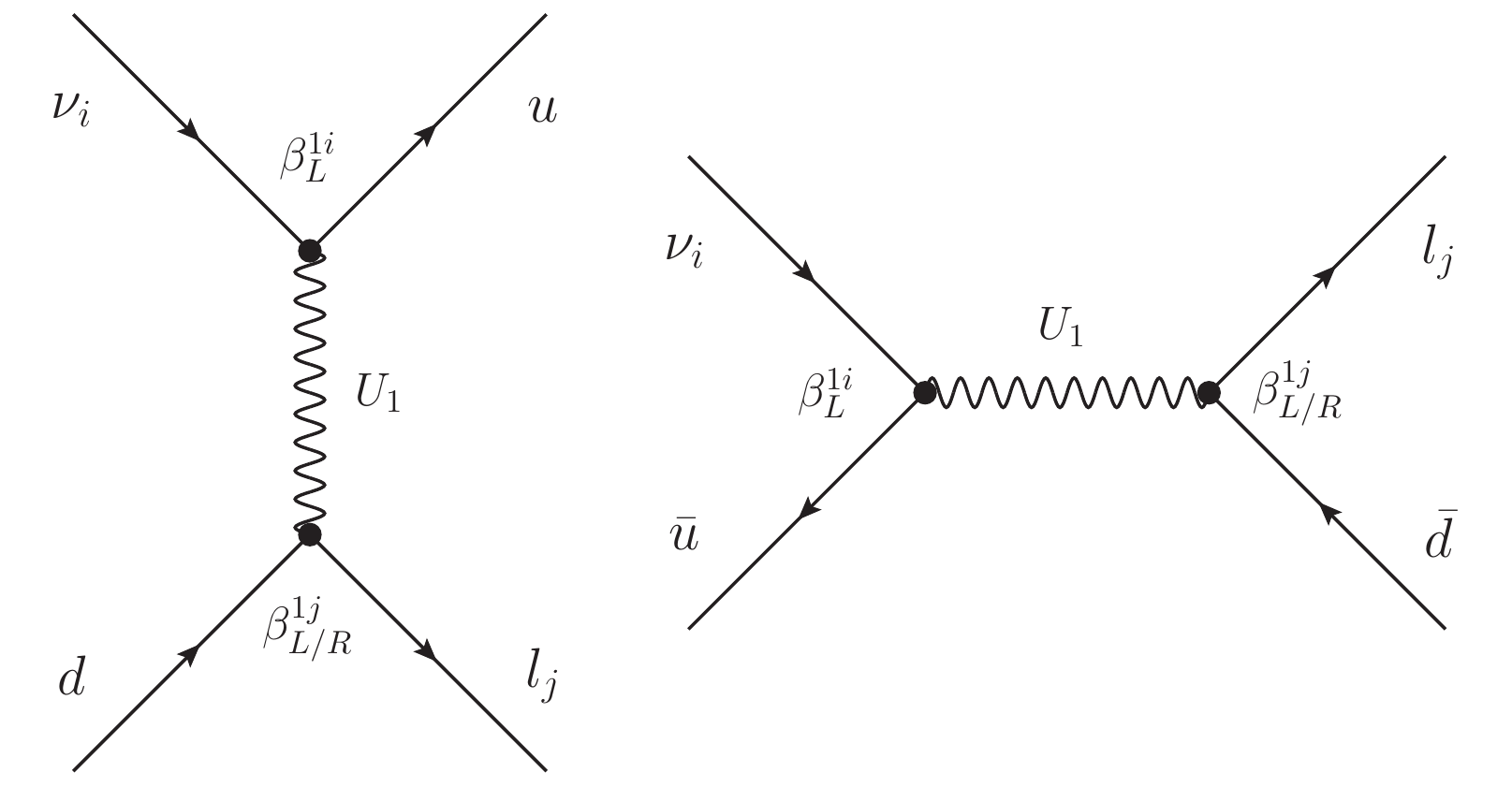}
\caption{Charged Current}
\end{subfigure}
\caption{ \label{fig1}
Contributing Feynman diagrams between the neutrino and the first-generation quark/antiquark through a vector-leptoquark exchange. The diagrams for antineutrinos and $d/\bar{d}$ particles can be similarly written down.
}
\end{figure}

In neutrino experiments, the leptoquark $U_1$ can participate in the neutral- and charged-current scattering between the incoming neutrino and the quark from protons or neutrons (nucleons) of the detector materials:
\begin{itemize}
\item Neutral current (NC): $\nu_{i} + N\longrightarrow \nu_{j} + N'$.
\item Charged current (CC): $\nu_{i}+N\longrightarrow l_{j} + N'$.
\end{itemize}
The contributing Feynman diagrams for the neutrino-quark scattering processes  are shown in Fig.~\ref{fig1}.
The amplitude for the NC scattering shown in Fig.~\ref{fig1} is given 
by, in the limit of heavy LQ mass, 
\begin{eqnarray}
i {\cal M}_{NC} &=& i \frac{g_U^2}{2 M^2_{U_1}} \beta_L^{1i} \beta_L^{1j}\, 
 (\overline{ u_L} \gamma^\mu \nu_{iL} ) \,
 (\overline{\nu_{jL}} \gamma_\mu u_{L} )  \nonumber \\
 &=& - i \frac{g_U^2}{2 M^2_{U_1} } \beta_L^{1i} \beta_L^{1j}\, 
 (\overline{u_L} \gamma^\mu u_{L} ) \,
 (\overline{\nu_{jL}} \gamma_\mu \nu_{iL} ) \;,
 \end{eqnarray}
 where the second line explicitly shows the NC scattering, which is 
 obtained by the Fierz transformation.
 On the other hand, the CC scattering for 
 $\nu_{iL} d_{L/R} \to u_L e_{L/R}$ can proceed via the left- and 
 right-handed couplings, of which the amplitude is given by, in the 
 limit of heavy LQ mass, 
 \begin{eqnarray}
  i {\cal M}_{CC} &=& i \frac{g_U^2}{2 M^2_{U_1}} \beta_L^{1i} 
  \, (\overline{u_L} \gamma^\mu \nu_{iL} ) \,
  \left( \beta_L^{1j}\, \overline{e_{jL}} \gamma_\mu d_L +
        \beta_R^{1j}\, \overline{e_{jR}} \gamma_\mu d_R \right ) \nonumber \\
        &=& i \frac{g_U^2}{2 M^2_{U_1} } \beta_L^{1i} \,
     \left[  - \beta_{L}^{1j} (\overline{u_L} \gamma^\mu d_{L} )\, 
                              (\overline{e_{jL}} \gamma_\mu \nu_{iL}  )
              + 2 \beta_R^{1j}(\overline{u_L}  d_{R} )\, 
                              (\overline{e_{jR}}\nu_{iL}  )
    \right ] \;,
    \end{eqnarray}
where the CC scattering is demonstrated explicitly in the second line,
which is obtained by the Fierz transformation.

Since we consider fixed-target scattering between neutrinos/antineutrinos and nucleon (detector), we neglect the effect of gluon-leptoquark interaction, and also the interaction between 
the leptoquark and photon or neutral $Z$ boson. 
Note that neutrinos will not scatter with electrons via the LQ $U_1$,
but, on the other hand, antineutrinos do scatter with electrons
via the LQ $U_1$. This can be distinguished from the neutrino-nucleon
scattering by the recoil nucleus. 


The proposed Forward Physics Facility (FPF) 
is set to be placed at several hundred meters from the ATLAS 
interaction point, shielded by concrete and rock \cite{Anchordoqui:2021ghd}. This FPF will 
house a number of experiments that will explore processes of the SM, 
as well as look for any physics beyond the Standard Model (BSM) 
\bibs{FASER:2019aik, FASER:2021ljd, FASER:2021cpr,FASER:2020gpr,FASER:2018bac, Ismail:2020yqc, Ansarifard:2021elw, Jodlowski:2020vhr, Feng:2017vli, Kling:2018wct, Feng:2018pew, Deppisch:2019kvs, Bakhti:2020szu, Jho:2020jfz, Okada:2020cue, Bahraminasr:2020ssz,Kelly:2020pcy,Falkowski:2021bkq,FASER:2018eoc, Asai:2022zxw, Cottin:2021lzz,Ismail:2021dyp,Batell:2021aja,Batell:2021snh,Cheung:2022kjd,Cheung:2022oji,Cheung:2021tmx,Aloni:2022ebm,Arakawa:2022rmp}. 
Such experiments are necessitated due to the high energy collisions at the High-Luminosity Large Hadron Collider (HL-LHC), which generates a large number of particles along the beam collision axis, beyond the scope of current LHC experiments.

A plethora of hadrons, such as pions, kaons, and more, are known to be produced along the beam direction. As these hadrons decay during the
flight, they produce neutrinos of all three flavors at very high energies up to a few TeV. Studies have revealed that muon neutrinos 
are primarily created from charged-pion decays, electron neutrinos arise from hyperon, kaon and $D$-meson decays, and tau neutrinos stem from $D_s$ meson decays. These neutrinos have an average energy ranging from 600 GeV to 1 TeV, comprising a wide energy range for each of the three neutrino flavors.

The following neutrino detectors are either operational and proposed for the far forward region of the LHC:

\begin{itemize}
\item \textcolor{blue}{FASER$\nu$}: A targeted mass located at the front of the FASER main detector in a narrow trench (illustrated in Figure 2 of Ref. \cite{FASER:2019dxq}), that is made from 1.2 tons of {\bf Tungsten} with the size of 25 cm $\times$ 30 cm $\times$ 1.1 m.
\item \textcolor{red}{FASER$\nu$2}: a detector, which is designed as a much large successor to FASER$\nu$, has a total volume of tungsten target is 50 cm $\times$ 50 cm $\times$ 8 m, so the total mass is 20 tons. A full description of the detectors and their requirements can be found in Ref. \cite{Feng:2022inv}.
\item {\bf FLArE}: a proposed liquid {\bf argon} time projection chamber (LArTPC) with an active volume of 10 tons (\textcolor{teal}{FLArE-10}) to 100 tons (\textcolor{orange}{FLArE-100}).
\end{itemize}

To estimate the number of events that occur inside these detectors, 
we calculate the scattering cross sections of neutrinos/antineutrinos
with the nucleus of the detector materials.  At nucleon level, the cross section is related to the neutrino/antineutrino parton 
cross section as:
\begin{equation}
\sigma_{\nu N}=\sum\limits_{i}\int\limits_{0}^{1}dx_{2}f(x_{2}, Q_{2})\sigma_{\nu q_{i}},
\end{equation}
where $f(x_{2}, Q_{2})$ is the parton distribution function (PDF) 
of $q_i$ inside the proton or neutron. This PDF depends on the 
momentum fraction $x_{2}$ and the factorization scale $Q_{2}$,
which we have chosen $Q_{2}=m_{Z}$.
We use the datasets from the LHAPDF library in Ref. \cite{Buckley:2014ana}. 

\begin{figure}[th!]
\centering
\captionsetup{justification=centering}
\begin{subfigure}{0.496\textwidth}
\centering
\includegraphics[width=\textwidth]{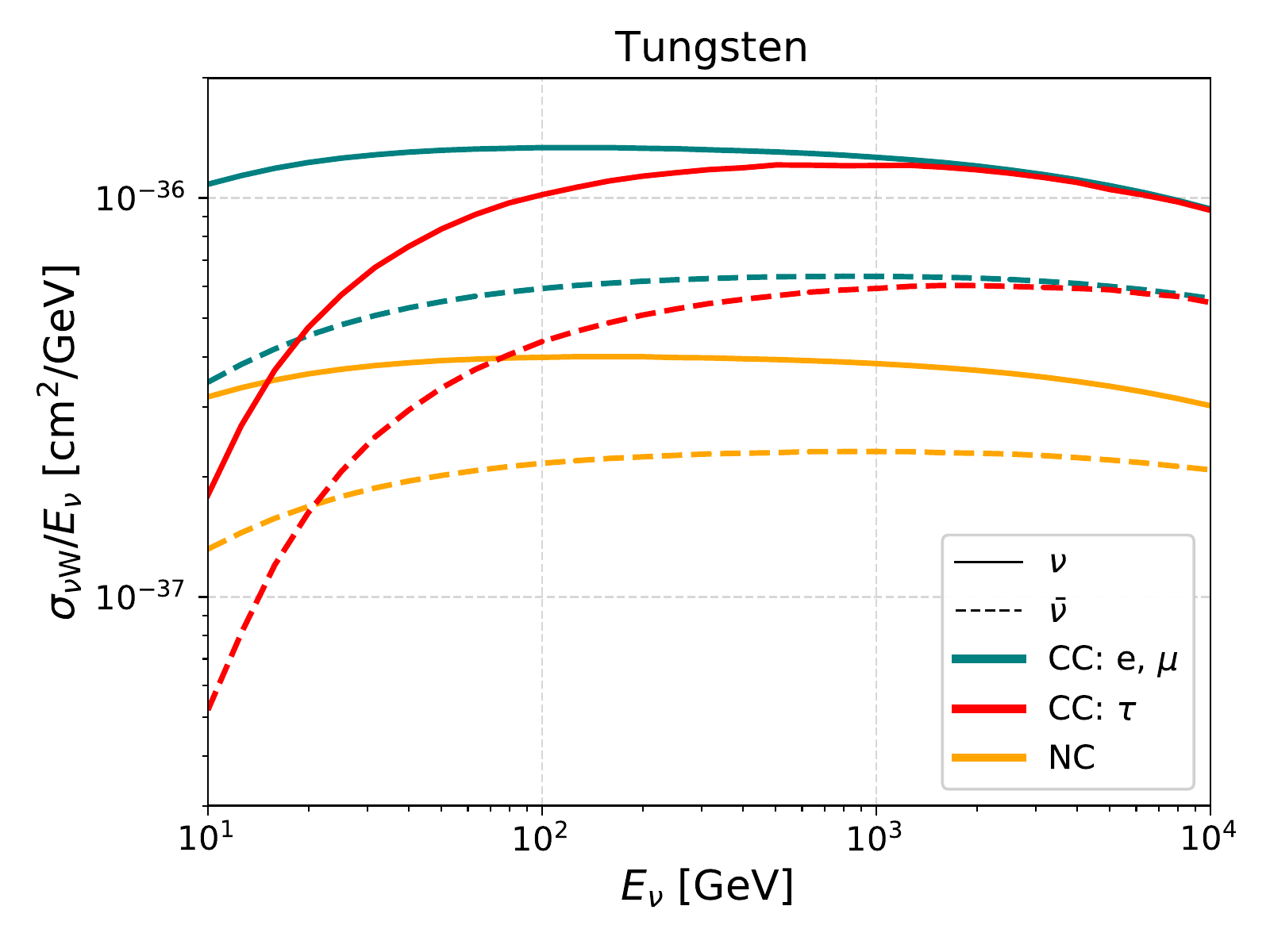}
\end{subfigure}
\hfill
\begin{subfigure}{0.496\textwidth}
\centering
\includegraphics[width=\textwidth]{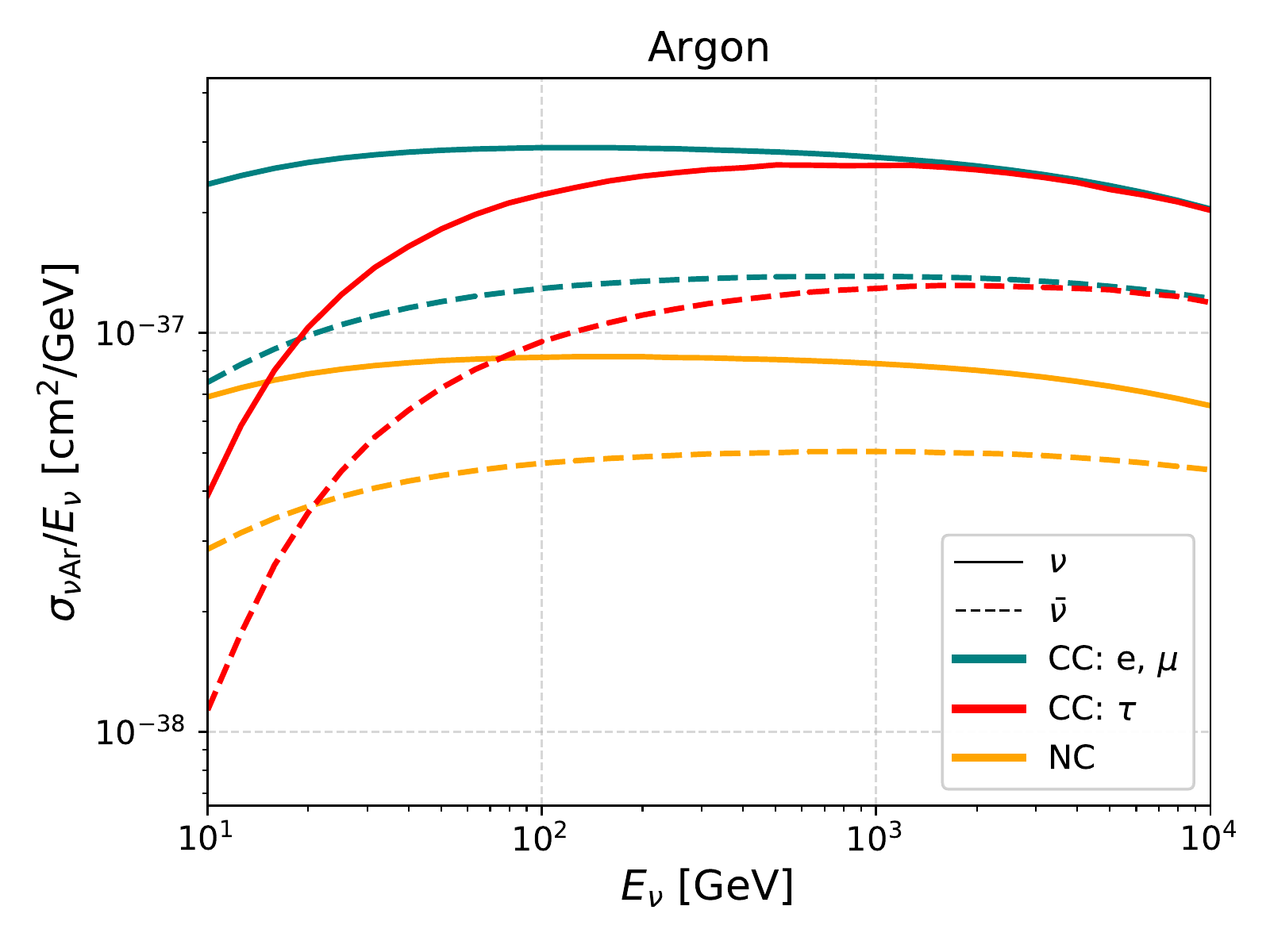}
\end{subfigure}
\caption{
The SM cross sections per neutrino energy for the scattering
of (anti)neutrinos with tungsten (left panel) and with argon (right
panel). Neutrino and antineutrino are separately shown. Neutral 
current (NC) and charged current (CC) for $(e,\mu)$ and $\tau$ are
shown.
}
\label{fig:SM_xsection}
\end{figure}

\begin{figure}[th!]
\centering
\captionsetup{justification=centering}
\begin{subfigure}{0.46\textwidth}
\centering
\includegraphics[width=\textwidth]{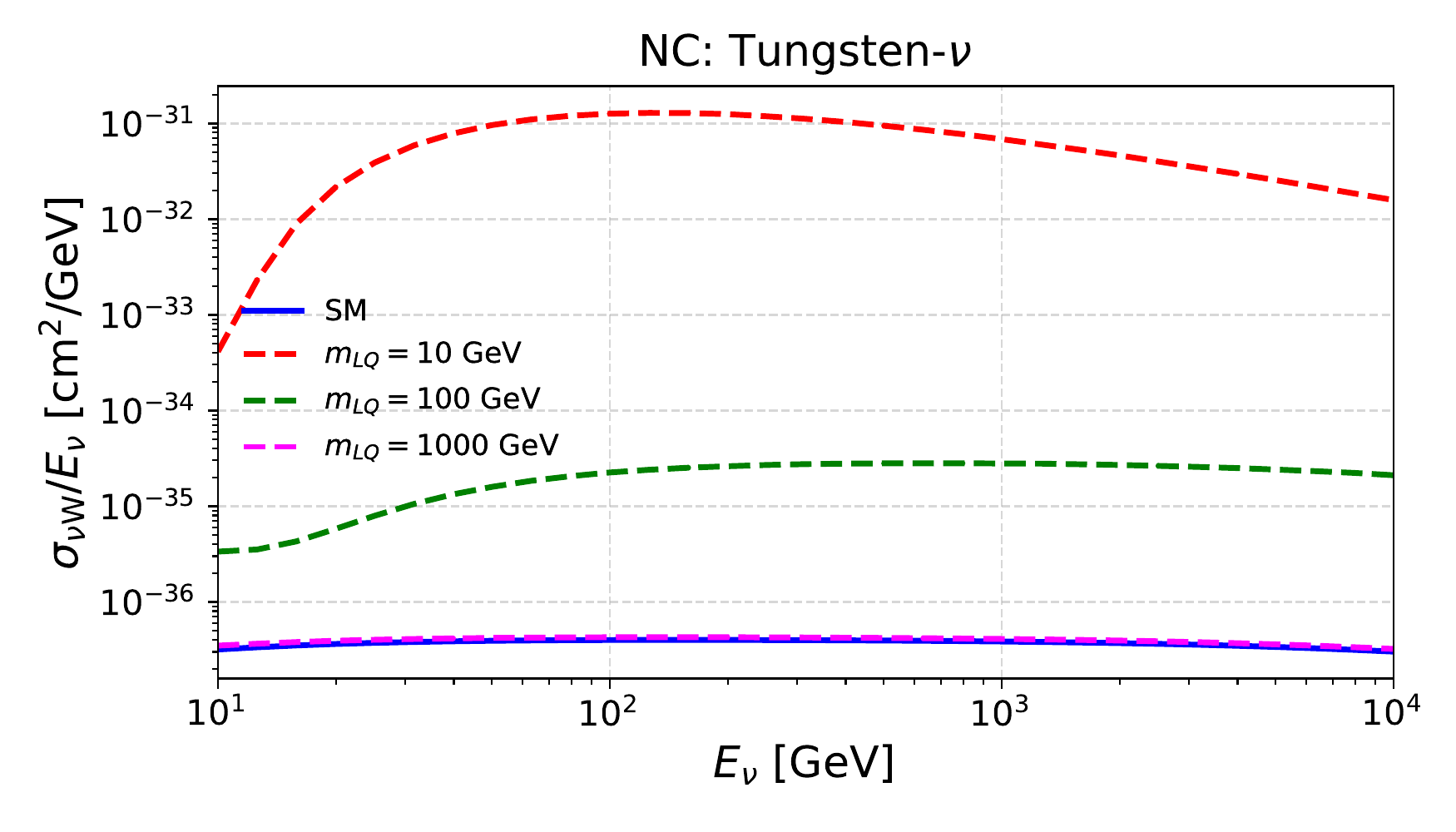}
\end{subfigure}
\hfill
\begin{subfigure}{0.46\textwidth}
\centering
\includegraphics[width=\textwidth]{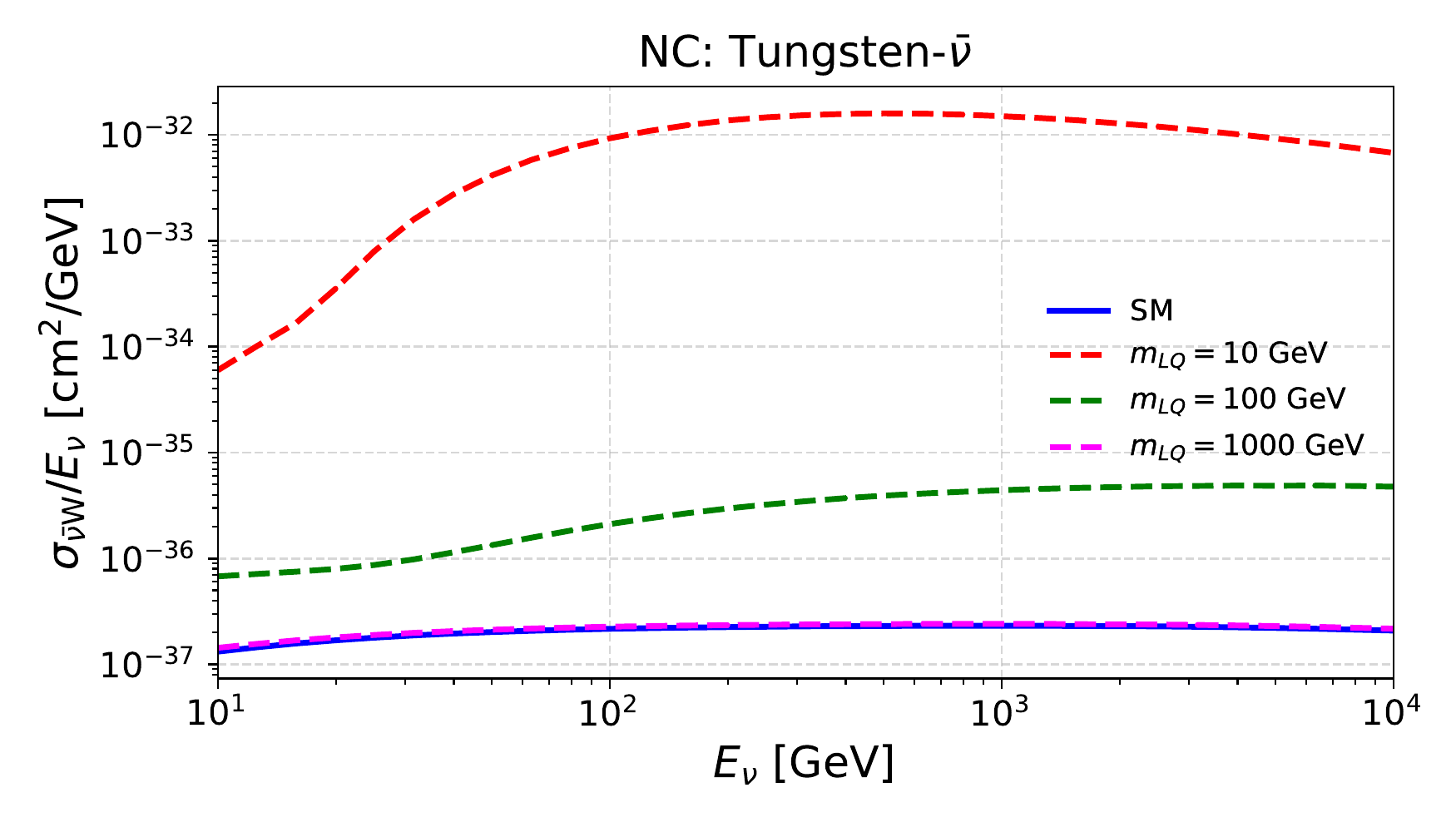}
\end{subfigure}
\begin{subfigure}{0.46\textwidth}
\centering
\includegraphics[width=\textwidth]{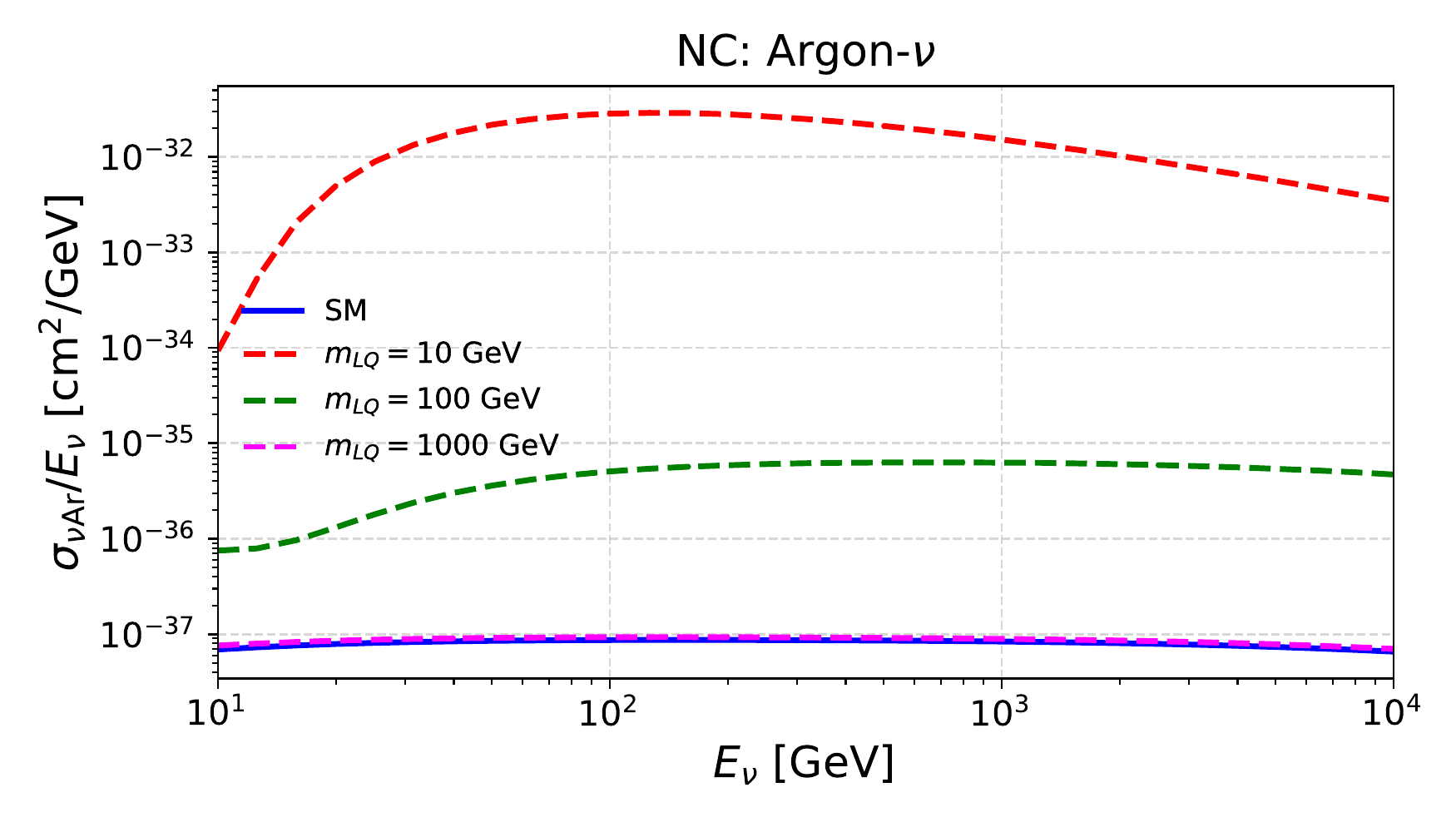}
\end{subfigure}
\hfill
\begin{subfigure}{0.46\textwidth}
\centering
\includegraphics[width=\textwidth]{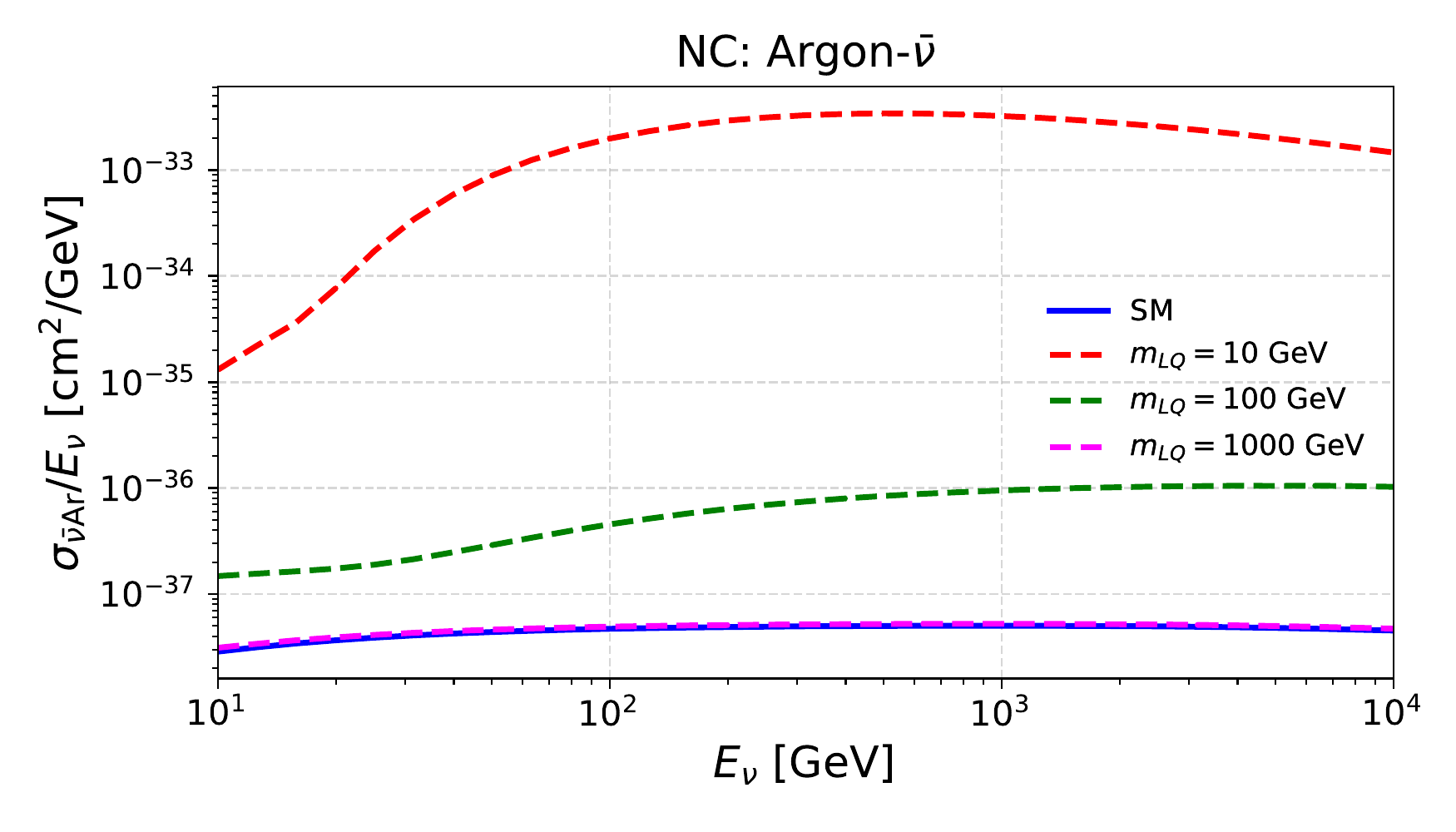}
\end{subfigure}
\caption{{\bf Neutral-Current (NC)} scattering cross sections per neutrino
energy versus the neutrino energy for both tungsten and argon 
targets.  We show the results for the SM and for various leptoquark
masses, and set all the couplings $g_{U}=\beta^{1j}=1$.} 
\label{fig:NC_xsection}
\end{figure}

\begin{figure}[thb!]
\centering
\captionsetup{justification=centering}
\begin{subfigure}{0.47\textwidth}
\centering
\includegraphics[width=\textwidth]{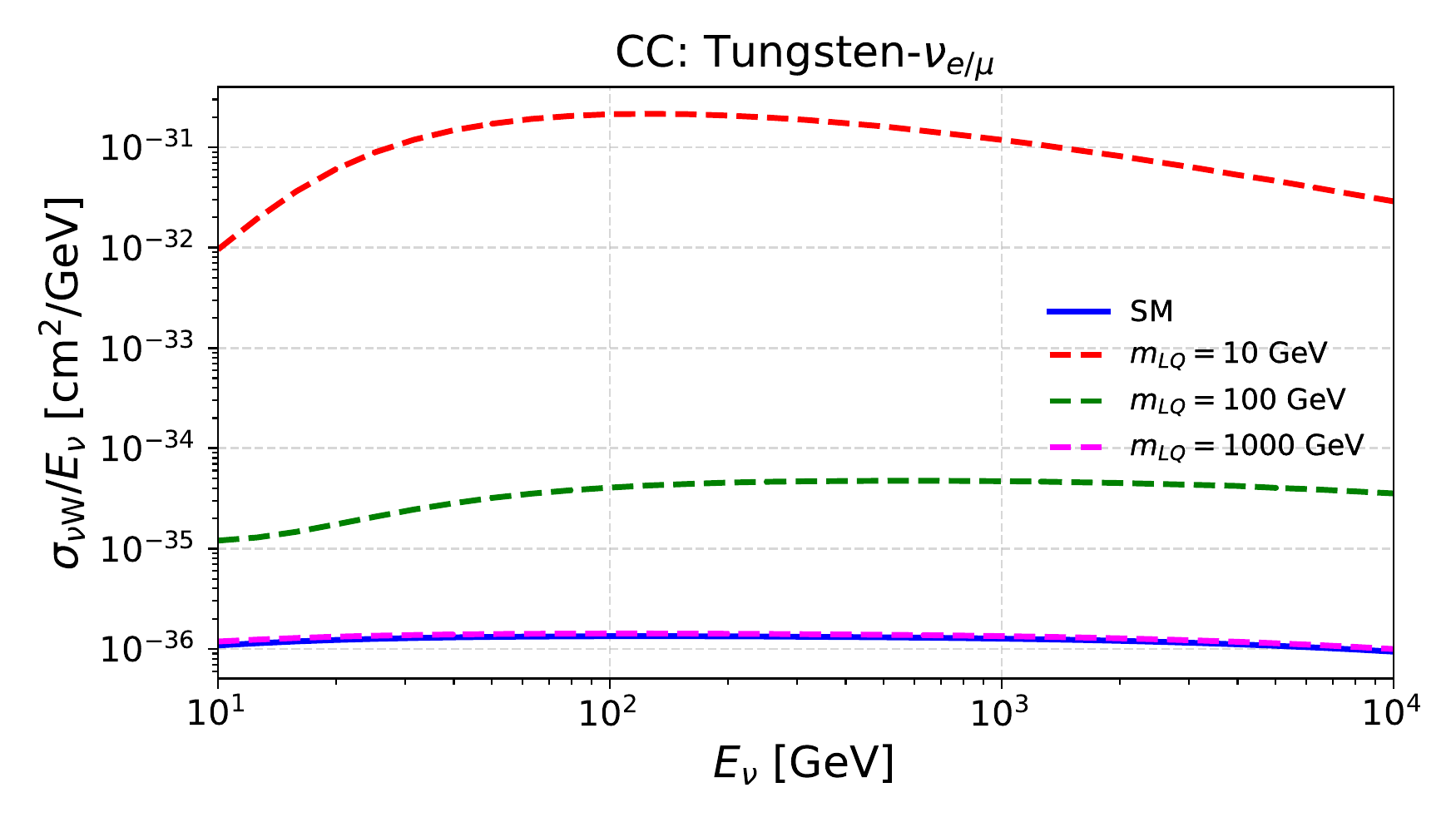}
\end{subfigure}
\hfill
\begin{subfigure}{0.47\textwidth}
\centering
\includegraphics[width=\textwidth]{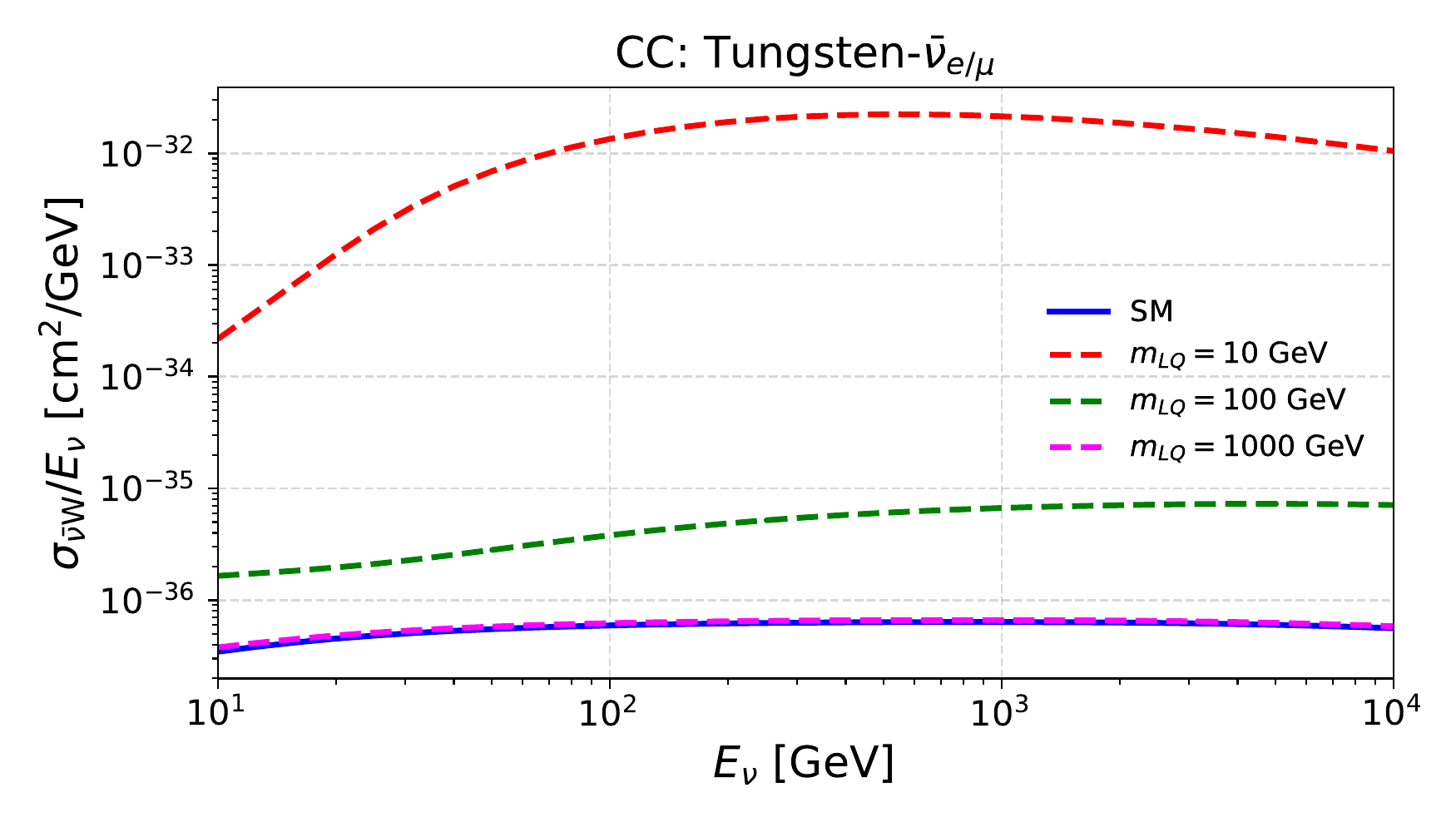}
\end{subfigure}
\begin{subfigure}{0.47\textwidth}
\centering
\includegraphics[width=\textwidth]{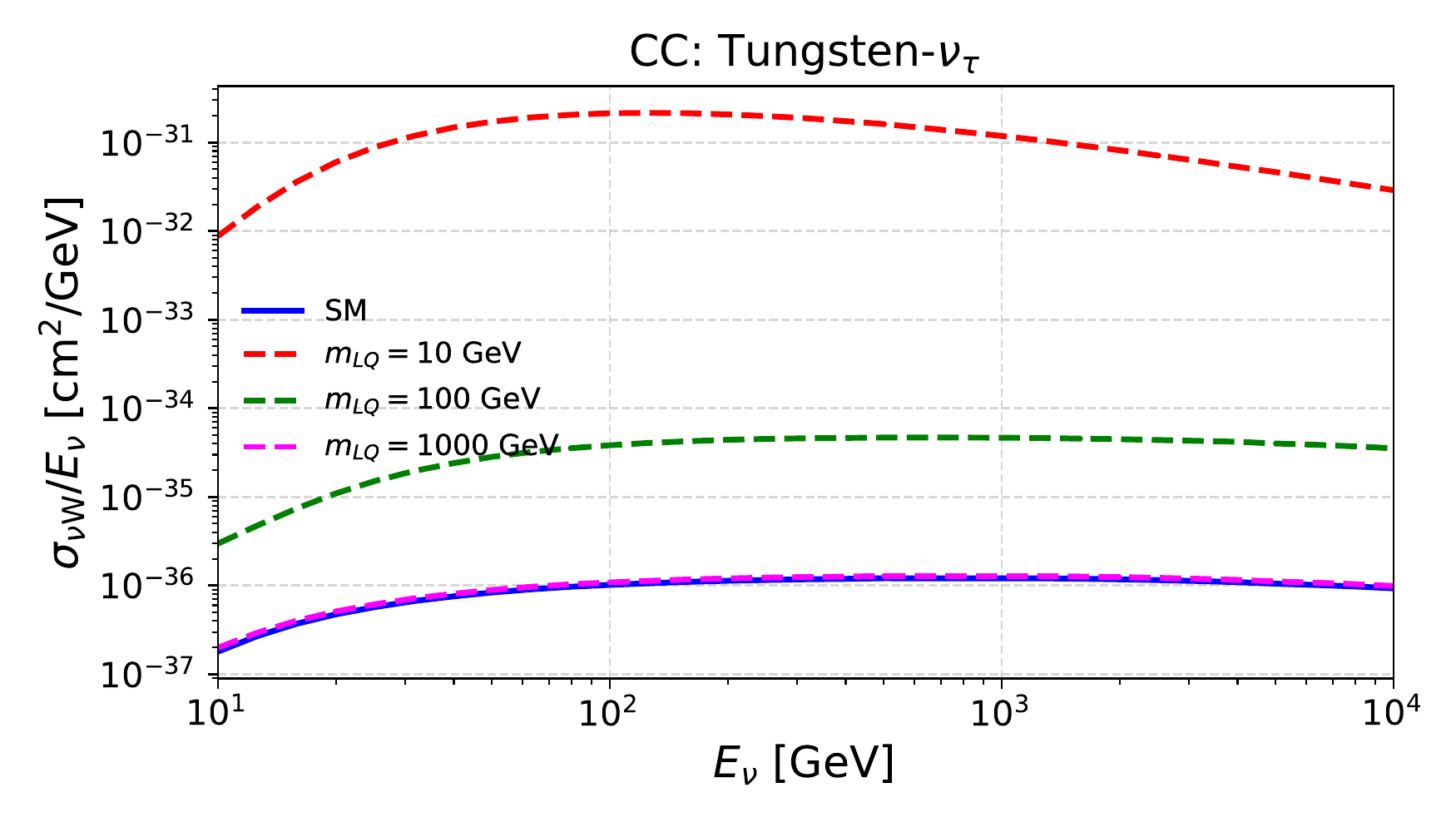}
\end{subfigure}
\hfill
\begin{subfigure}{0.47\textwidth}
\centering
\includegraphics[width=\textwidth]{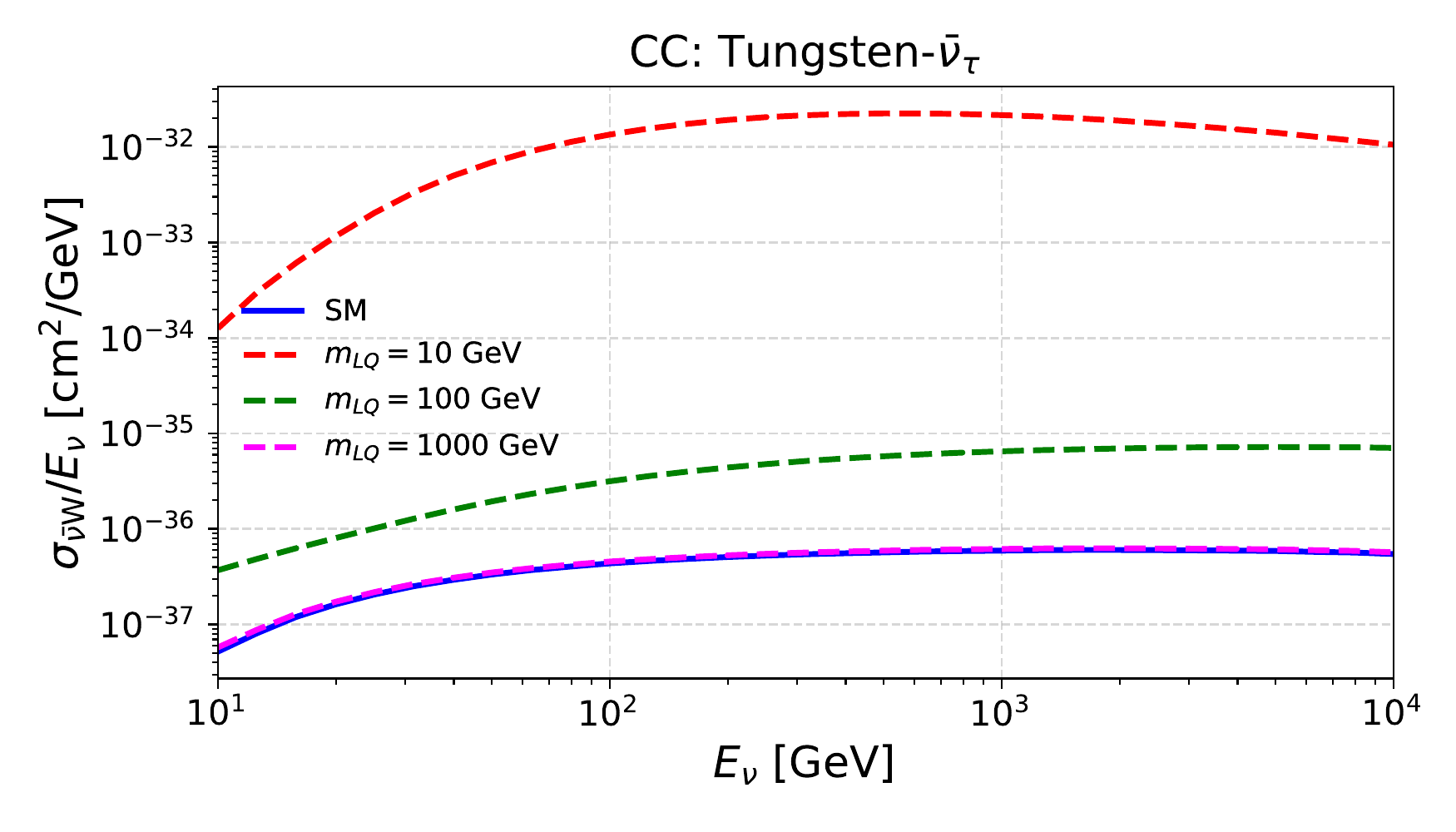}
\end{subfigure}
\begin{subfigure}{0.47\textwidth}
\centering
\includegraphics[width=\textwidth]{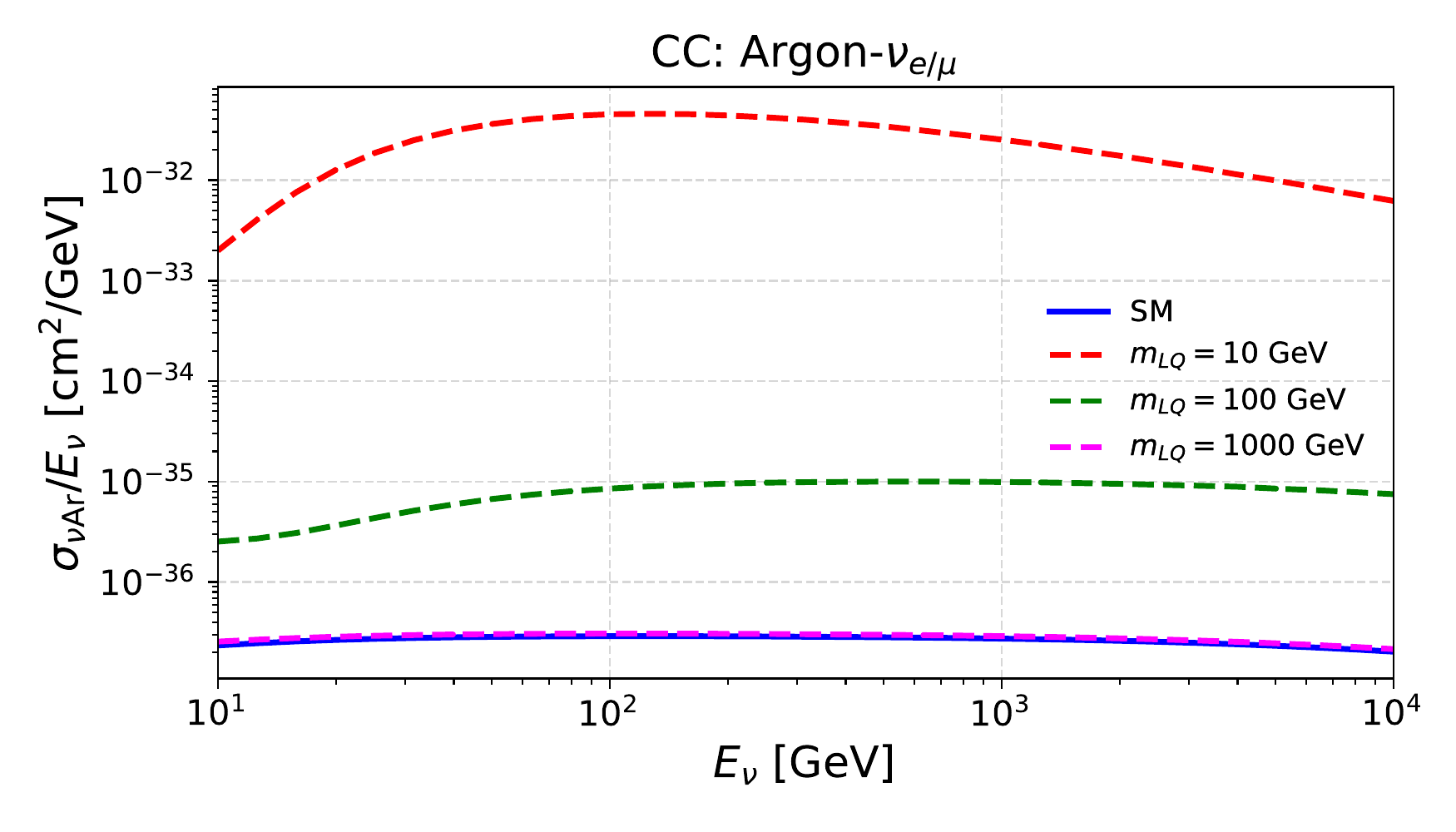}
\end{subfigure}
\hfill
\begin{subfigure}{0.47\textwidth}
\centering
\includegraphics[width=\textwidth]{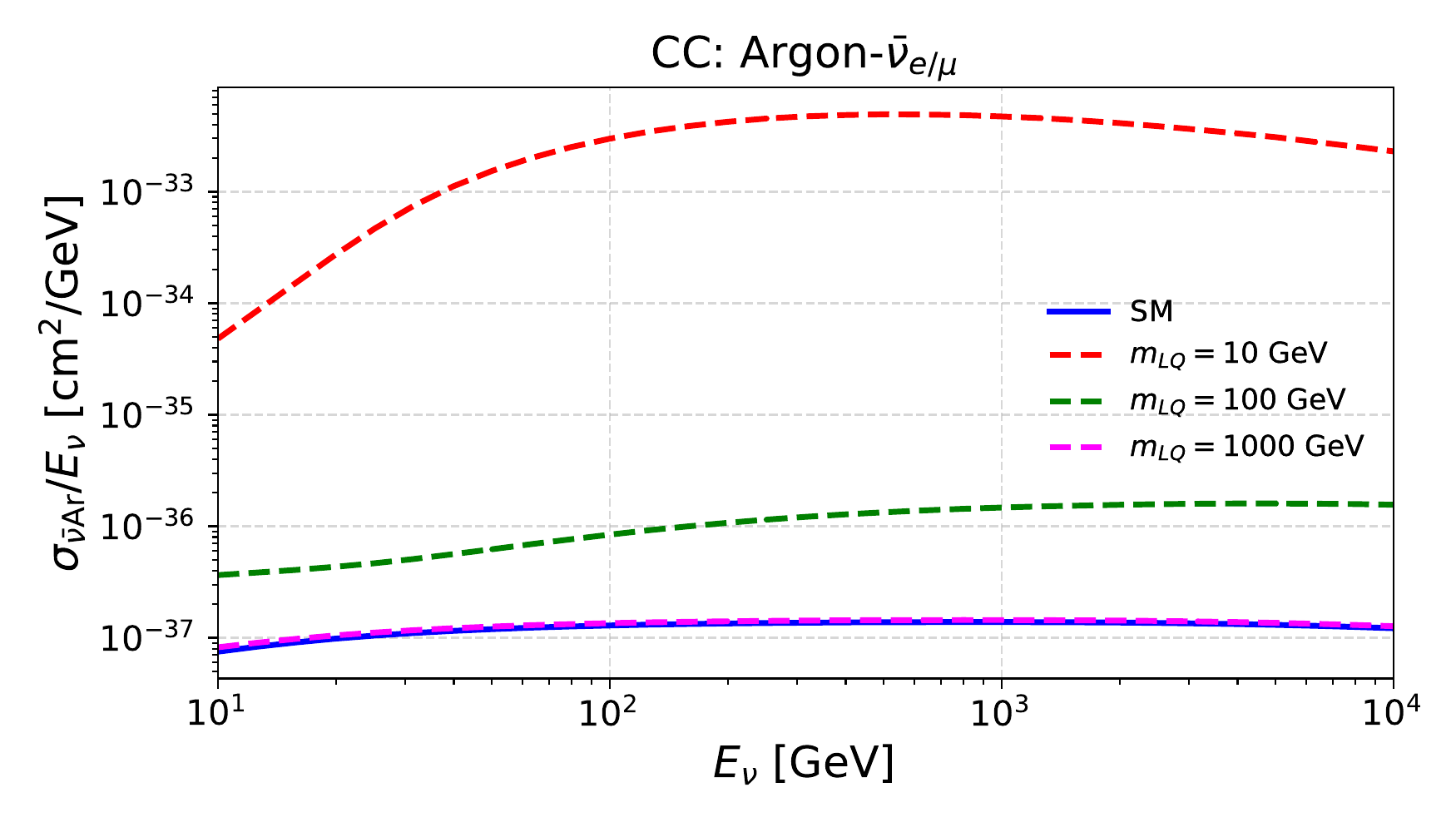}
\end{subfigure}
\hfill
\begin{subfigure}{0.47\textwidth}
\centering
\includegraphics[width=\textwidth]{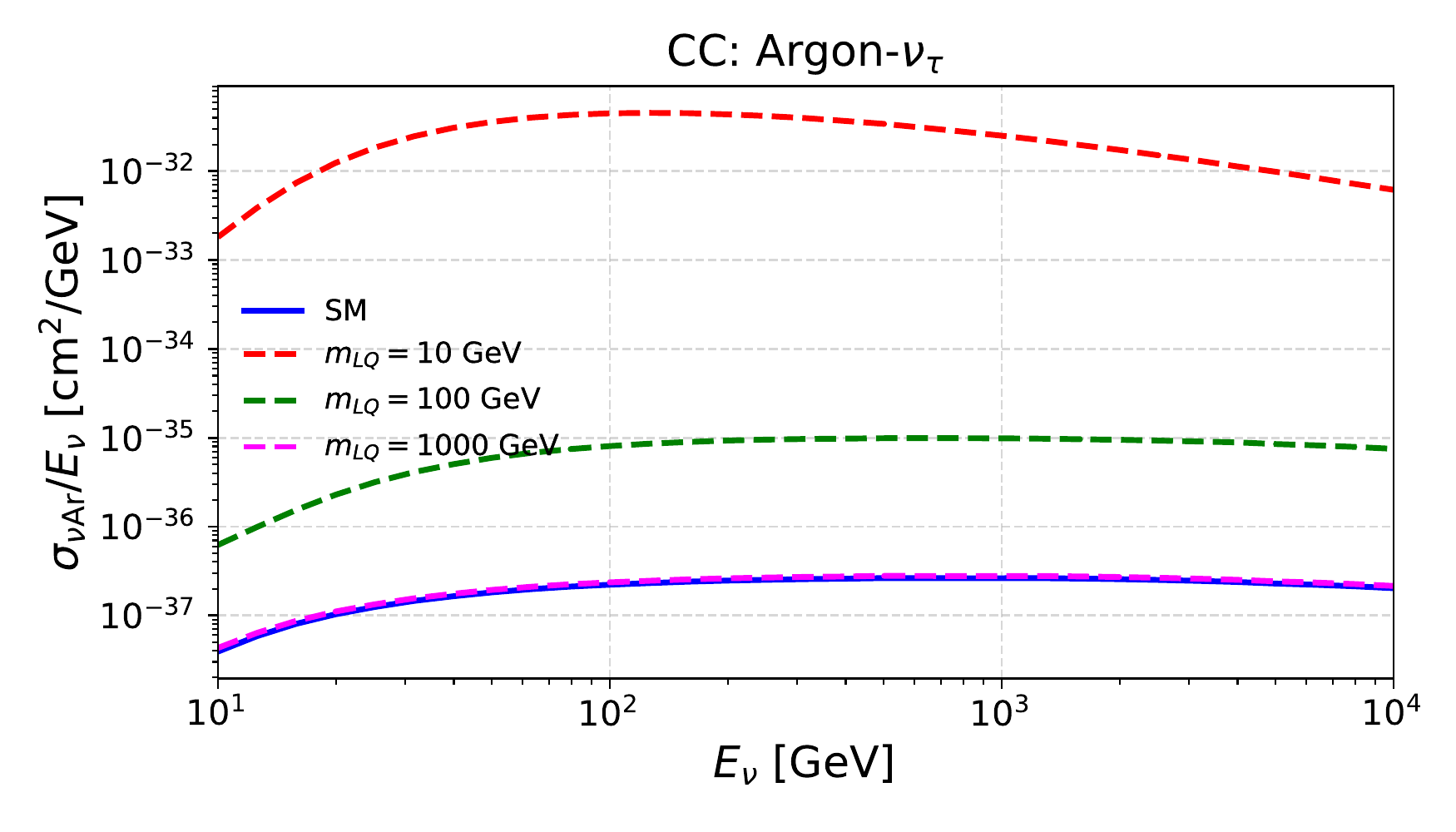}
\end{subfigure}
\hfill
\begin{subfigure}{0.47\textwidth}
\centering
\includegraphics[width=\textwidth]{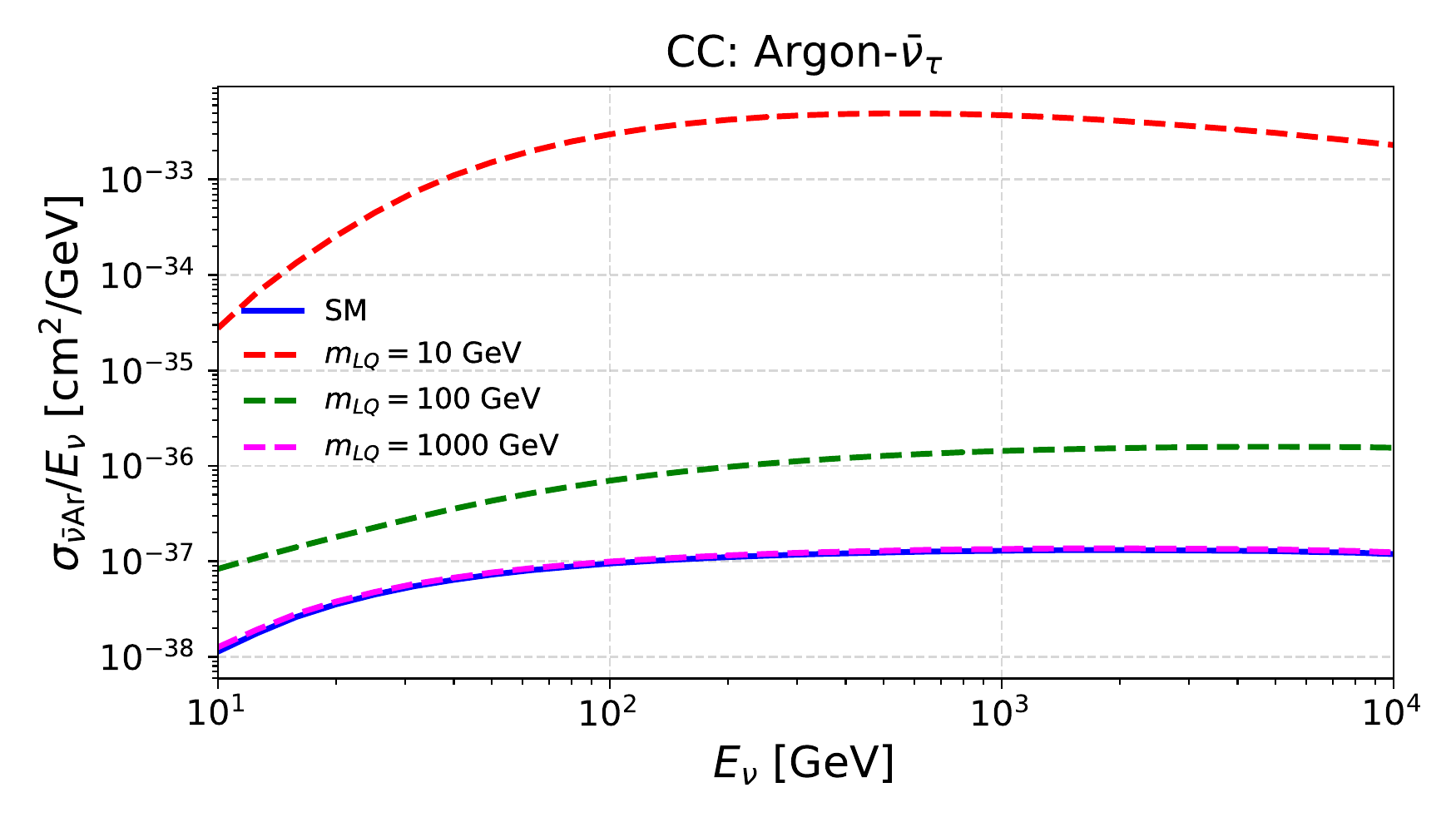}
\end{subfigure}
\caption{{\bf Charged Current (CC)} scattering cross sections per 
neutrino energy. All the details are similar to  
\fig{fig:NC_xsection}.} 
\label{fig:CC_xsection}
\end{figure}

We show the SM NC and CC cross sections for neutrinos and 
antineutrinos in each detector's material in \fig{fig:SM_xsection}.
In general, the neutrino gives a larger cross section than the 
anti-neutrino,
and the CC cross section is larger than the NC one. Note that
the CC tau-neutrino and tau-anti-neutrino have lower cross sections
than corresponding ones of electron and muon, because of a higher 
threshold to produce a tau lepton. 
The effects of leptoquark interactions are shown for NC and CC 
scattering in \fig{fig:NC_xsection} and \fig{fig:CC_xsection}, 
respectively, for various leptoquark masses $10-1000$ GeV.
Here we have set $g_U$ and $\beta^{1j}_{L/R} =1$. 

To calculate the CC and NC cross sections for each process, we use 
the $U_{1}$ leptoquark model file, which is a modified version of \refs{DiLuzio:2018zxy, Baker:2019sli, Cornella:2021sby}, and 
input it to FeynRules to generate a model file for use in 
$\rm MadGraph5aMC@NLO$ (\refs{Alwall:2014hca, Alwall:2011uj}). 
To match the results from \bib{FASER:2019dxq}, we used
the same NNPDF3.1NNLO PDF (\bib{NNPDF:2017mvq}).



\begin{figure}[thb!]
\centering
\captionsetup{justification=centering}
\includegraphics[scale=0.5]{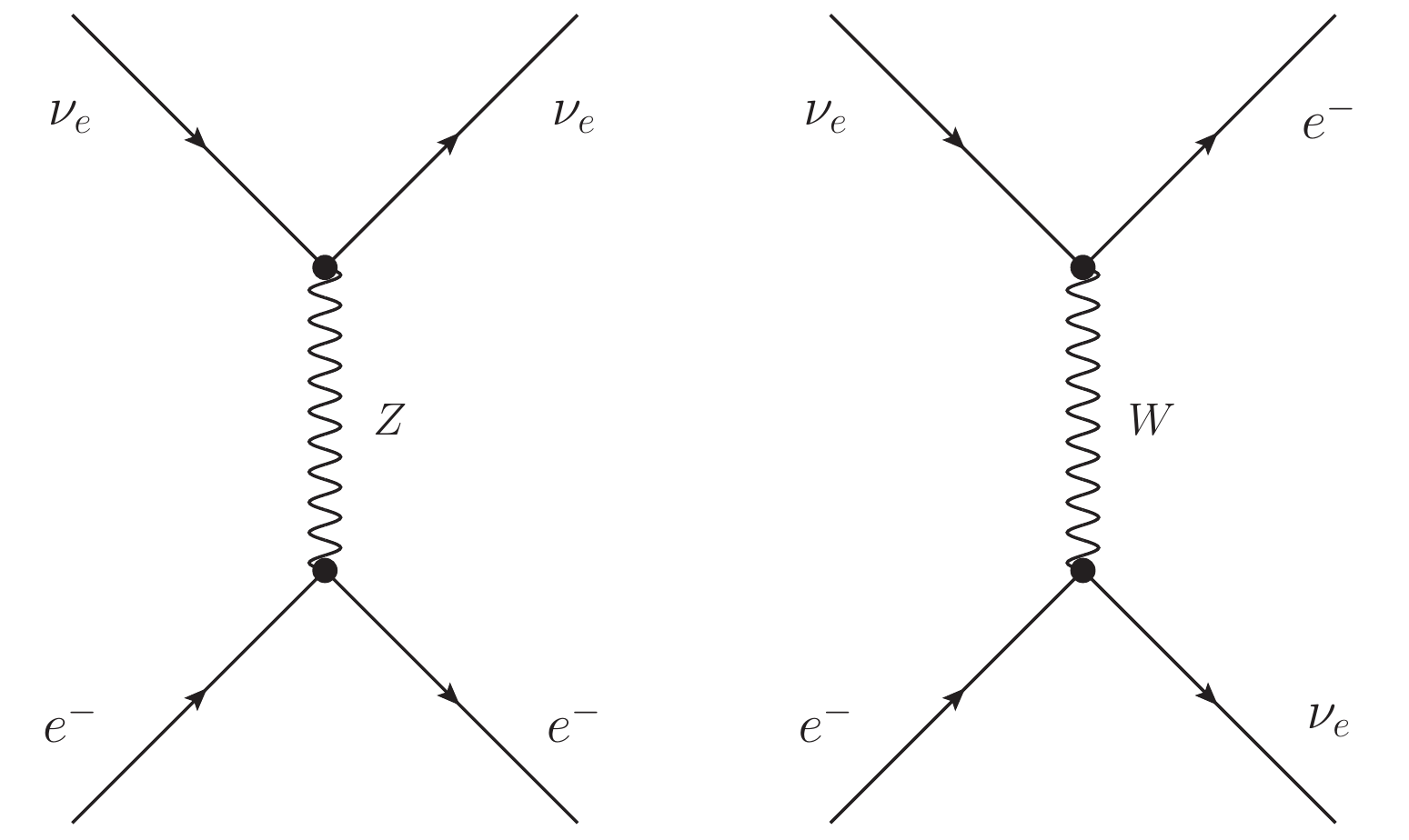}
\caption{Feynman diagrams for the neutrino-electron in 
tungsten/argon detectors, as the SM background for our process.}
\label{fig:SMBG}
\end{figure}

A discussion on the background coming from neutrino-electron
or antineutrino-electron scattering is in order here. We found that
the leptoquark exchange is not possible in the neutrino-electron
scattering, while it is possible to participate in 
antineutrino-electron scattering $\bar \nu e^- \to \bar u d$
via the $t$-channel LQ exchange. Nevertheless, the nucleus remains
intact in this case. We can make use of the recoiled nucleus to
distinguish the signal from this background.
However, the hadronic part in the CC $\nu$-nucleon scattereing 
may not be fully identified with a 100\% efficiency. 
Therefore, whether this reducible background can be completely 
removed depends on the experimental setup and resolution. 
We take into account this background uncertainty  
by including it into the 
systematic uncertainty in the estimation of sensitivities.

Besides cross sections, we have to take into account the 
neutrino and antineutrino fluxes to calculate the number of 
events. These fluxes are measured and distinguishable for 
each  (anti)neutrino flavor in FASER$\nu$, FASER$\nu$2, 
and FLArE detectors. We used the fluxes calculated 
in \bib{FASER:2019dxq} for FASER$\nu$, and the simulated 
fluxes for FASER$\nu$2, FLArE 10 and 100 tons 
from \bib{Feng:2022inv}. The uncertainties in the incoming
(anti)neutrino fluxes were studied in \bib{Kling:2021gos}. 
The electron-neutrino flux has less than 10\% uncertainty up to 
0.8 TeV, while the dominated muon neutrino flux has less 
than 10\% uncertainty up to 1 TeV, and the tau-neutrino 
flux has less than 10\% uncertainty up to 0.3$\sim$0.4 TeV. 
Overall, the uncertainties are within a factor of 2. 
Such uncertainties will propagate to the event rate predictions 
of order a few percent up to 20\%, so we can include them in the systematic uncertainties. We take into account the uncertainties due to the hadronic efficiencies
and the incoming neutrino flux by including them into the systematic 
uncertainties of order 10\% of the SM predictions \cite{Feng:2022inv}, 
which we expect to go down with more intensive studies.

\section{Sensitivity on the Vector Leptoquark $U_1$}

\begin{figure}[thb!]
\centering
\captionsetup{justification=centering}
\begin{subfigure}{0.47\textwidth}
\centering
\includegraphics[width=\textwidth]{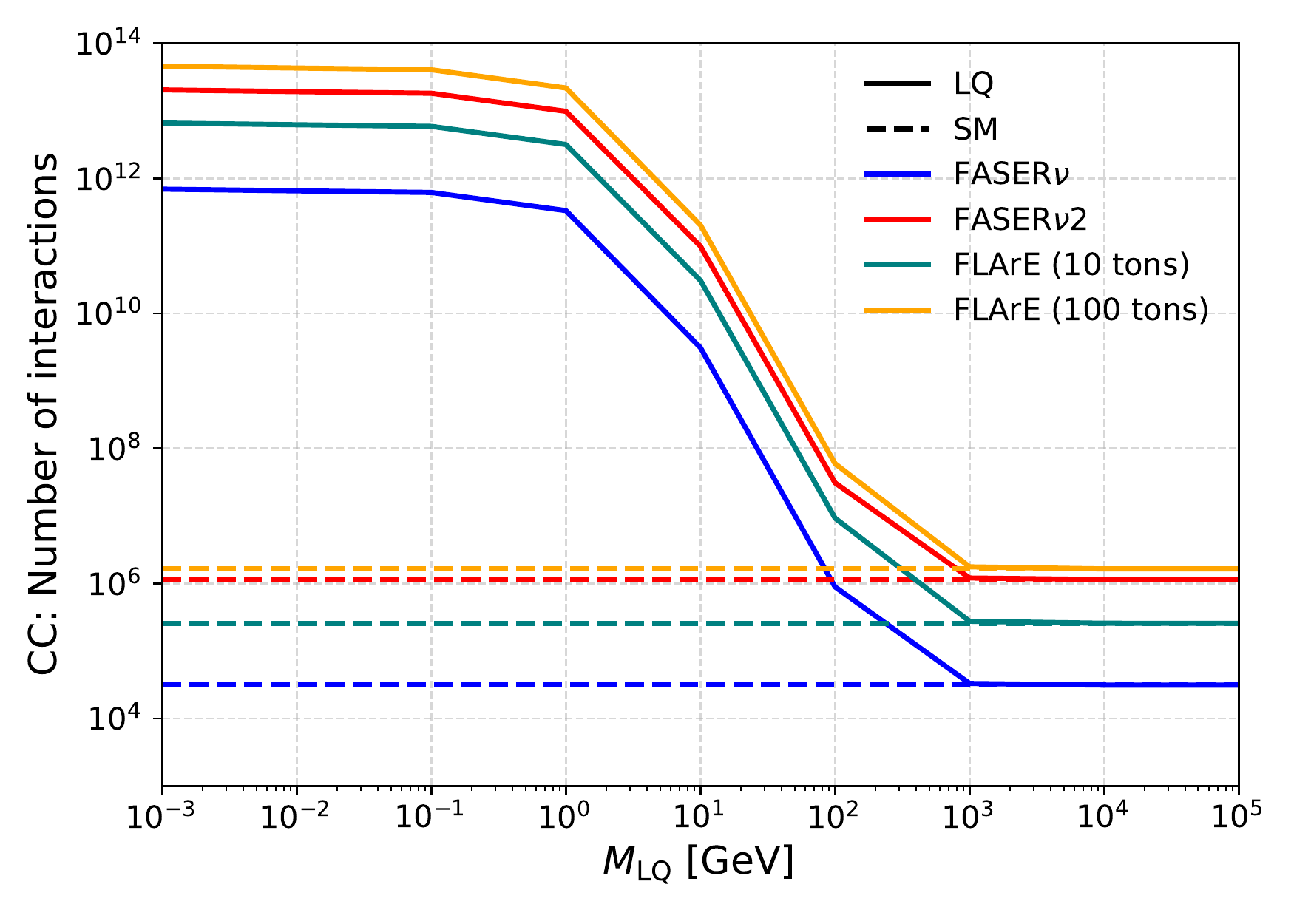}
\end{subfigure}
\hfill
\begin{subfigure}{0.47\textwidth}
\centering
\includegraphics[width=\textwidth]{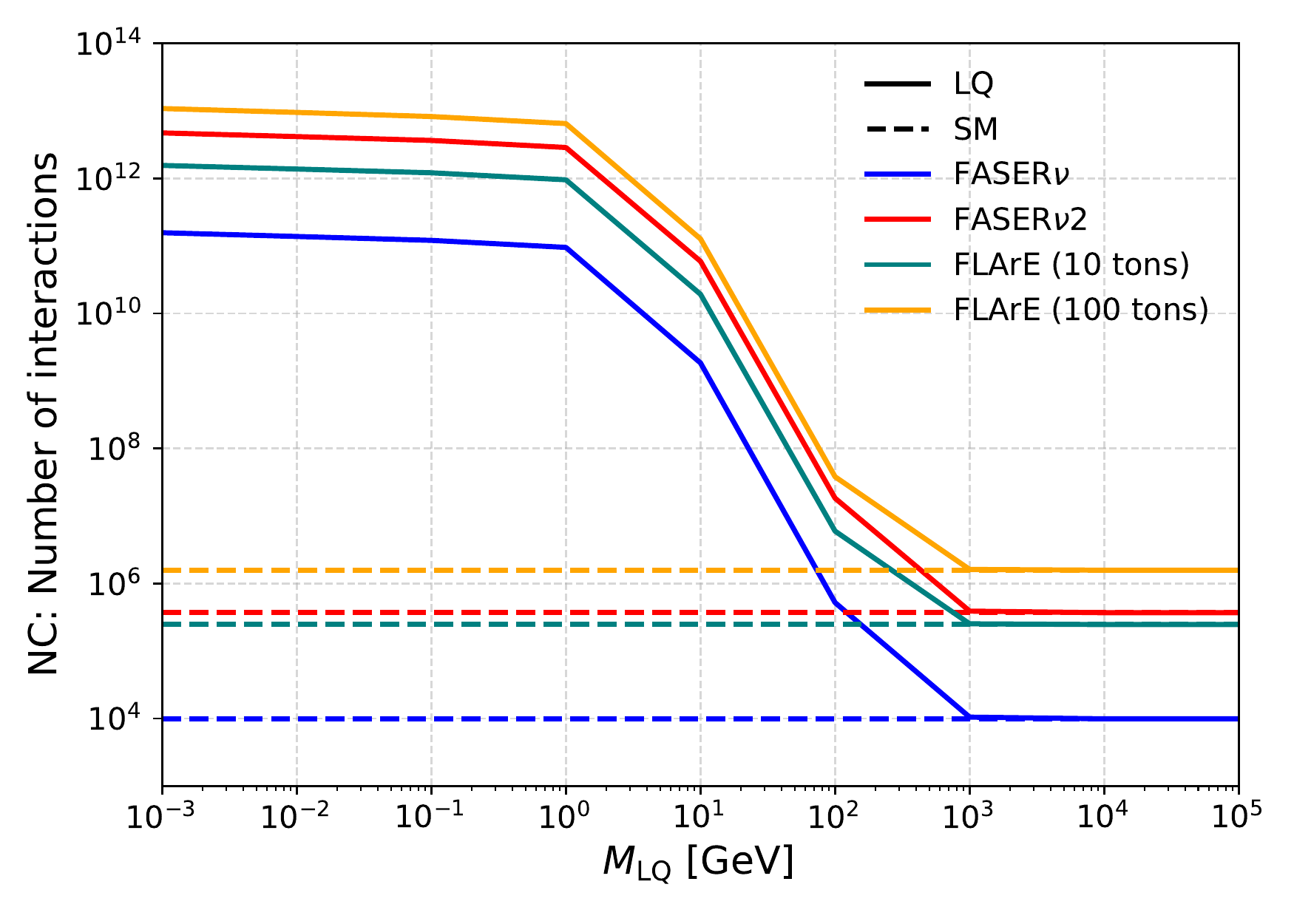}
\end{subfigure}
\hfill
\caption{Number of events for the FPF detectors versus the leptoquark
mass. We set all the couplings $g_{U}=\beta^{1j}=1.0$. Here the solid
lines are for the leptoquark while the dashed lines are for the SM.}
\label{fig:Nevents}
\end{figure}

Using the neutrino fluxes \cite{FASER:2019dxq,Feng:2022inv} 
and the cross sections calculated for both SM and the leptoquark model,
we show the total number of events for various FPF detectors 
in Fig. \ref{fig:Nevents}. 
%
Once we obtain the predicted number of events $N_{\mathrm{BSM}}$ for
the leptoquark model as a function of the coupling, and that for
the SM $N_{\mathrm{SM}}$, we estimate the sensitivity reach in the
parameter space ($g_{U}$ or $g_{U}\times \beta^{1j}$ vs $M_{LQ}$) of 
the model. 
Here we have taken the statistical error to be 
$\sqrt{N_{\mathrm{BSM}}}$ and the systematic uncertainty to be  
$\sigma=10\%$ of the normalization of the SM predictions. 
For all three generations of neutrinos, we defined the measure of $\chi^{2}$ as a function of the coupling $g_{\mathrm{BSM}}=g_{U}$ or
$g_{U}\times \beta^{1j}$ and a nuisance parameter $\alpha$ as follows \bib{Ballett:2019xoj}

\begin{equation}
\begin{split}
\chi^{2}(g_{\mathrm{BSM}}, \alpha)=\min\limits_{\alpha}\Big{[}&\frac{[N^{\nu_{e}}_{\mathrm{BSM}}-(1+\alpha)N^{\nu_{e}}_{\mathrm{SM}}]^{2}}{N^{\nu_{e}}_{\mathrm{BSM}}}+\frac{[N^{\nu_{\mu}}_{\mathrm{BSM}}-(1+\alpha)N^{\nu_{\mu}}_{\mathrm{SM}}]^{2}}{N^{\nu_{\mu}}_{\mathrm{BSM}}}\\
+&\frac{[N^{\nu_{\tau}}_{\mathrm{BSM}}-(1+\alpha)N^{\nu_{\tau}}_{\mathrm{SM}}]^{2}}{N^{\nu_{\tau}}_{\mathrm{BSM}}}+\Big{(}\frac{\alpha}{\sigma}\Big{)}^{2}\Big{]},
\end{split}
\end{equation}
where $N_{\mathrm{BSM}}=N_{LQ}+N_{\mathrm{Int}}+N_{\mathrm{SM}}$ and the minimization is over the nuisance parameter $\alpha$. 
Here $N_{LQ}$ is the number of events that only comes from 
leptoquark contribution, while $N_{\mathrm{Int}}$ is the interference term between the leptoquark and the SM. 
We have treated the systematic uncertainties the same for 
each neutrino flavor and used only one nuisance parameter $\alpha$. 
For the following analysis results, we choose $\chi^{2}=3.84$ for the 
95\% C.L. sensitivity reach.

\begin{figure}[thb!]
\centering
\captionsetup{justification=centering}
\begin{subfigure}{0.325\textwidth}
\centering
\includegraphics[width=\textwidth]{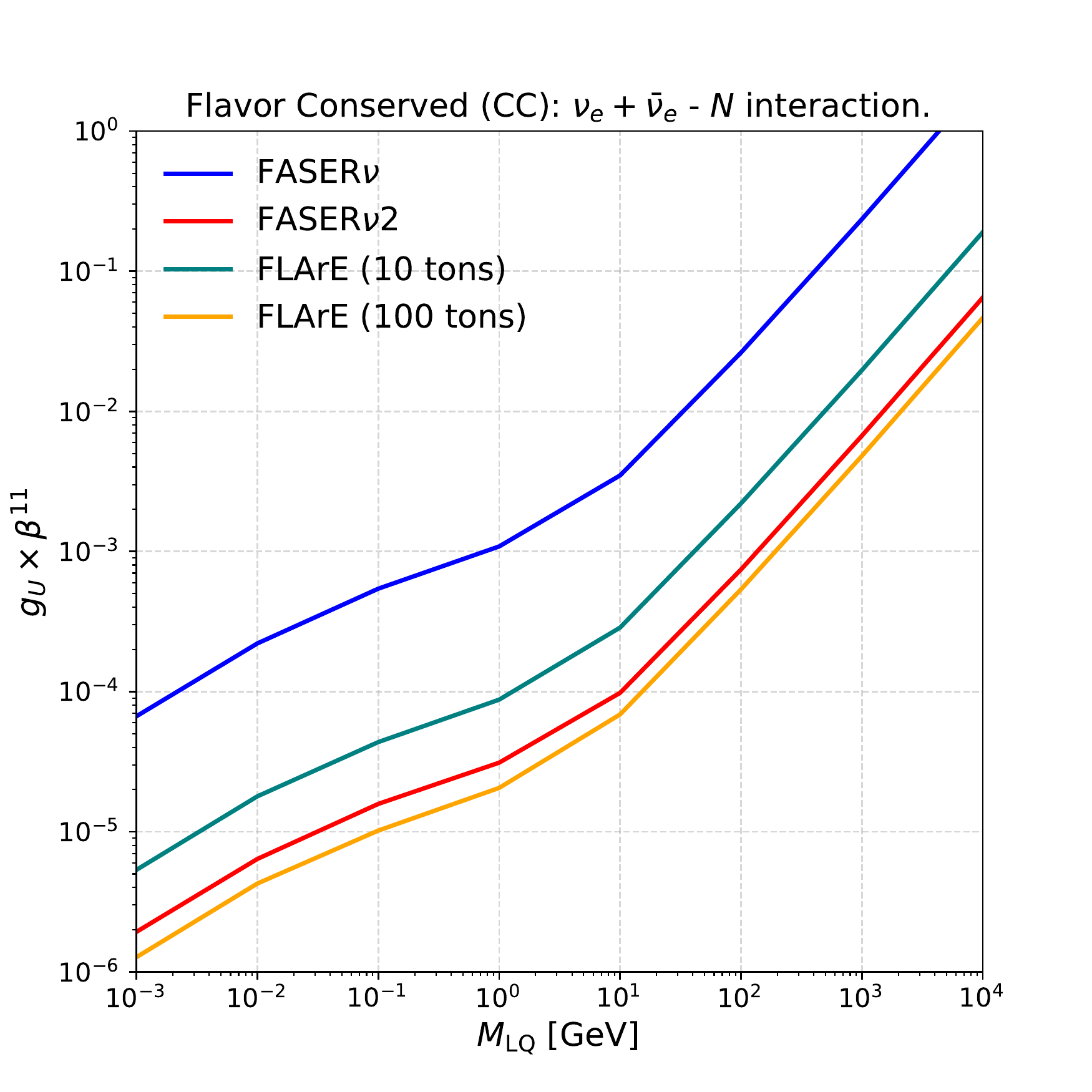}
\end{subfigure}
\hfill
\begin{subfigure}{0.325\textwidth}
\centering
\includegraphics[width=\textwidth]{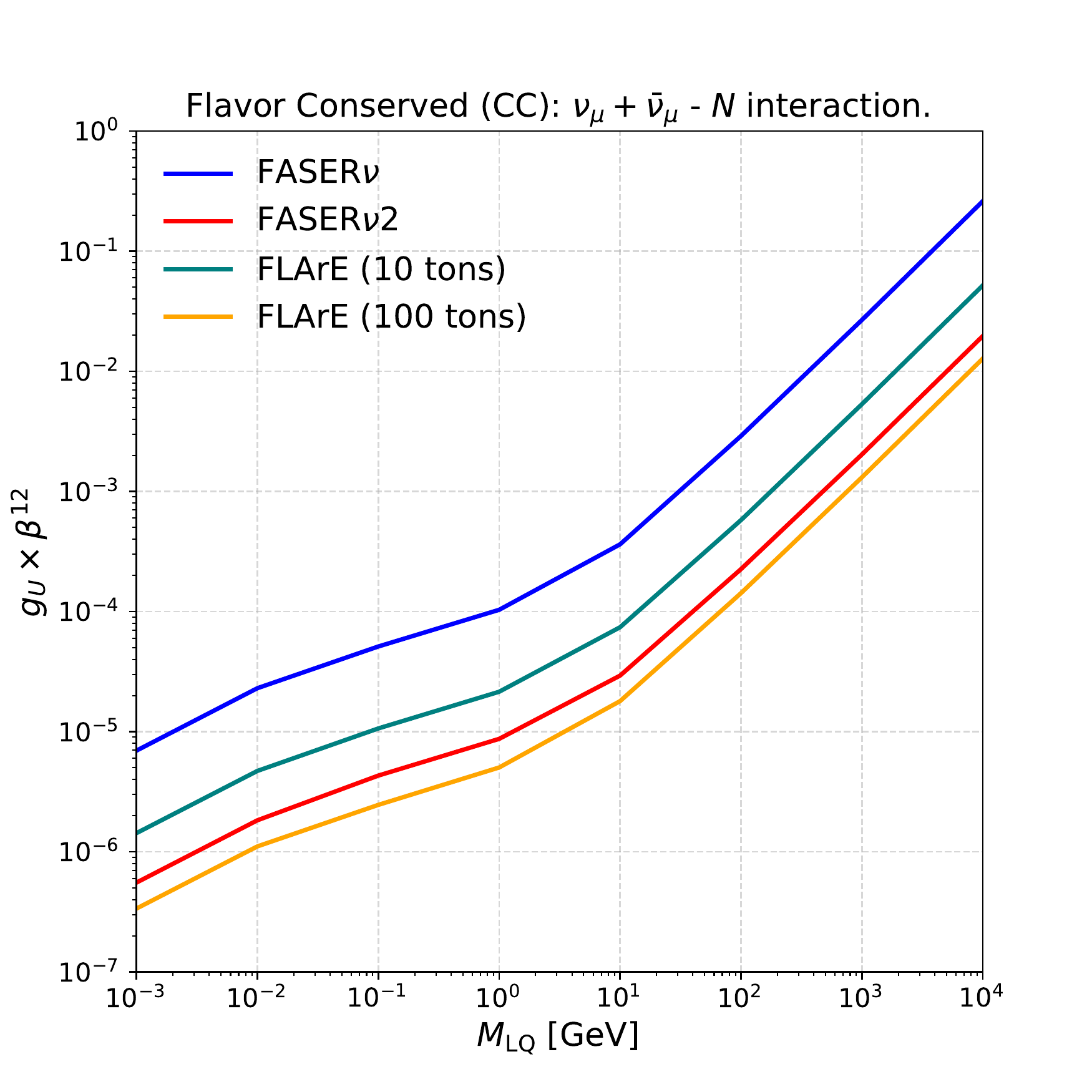}
\end{subfigure}
\hfill
\begin{subfigure}{0.325\textwidth}
\centering
\includegraphics[width=\textwidth]{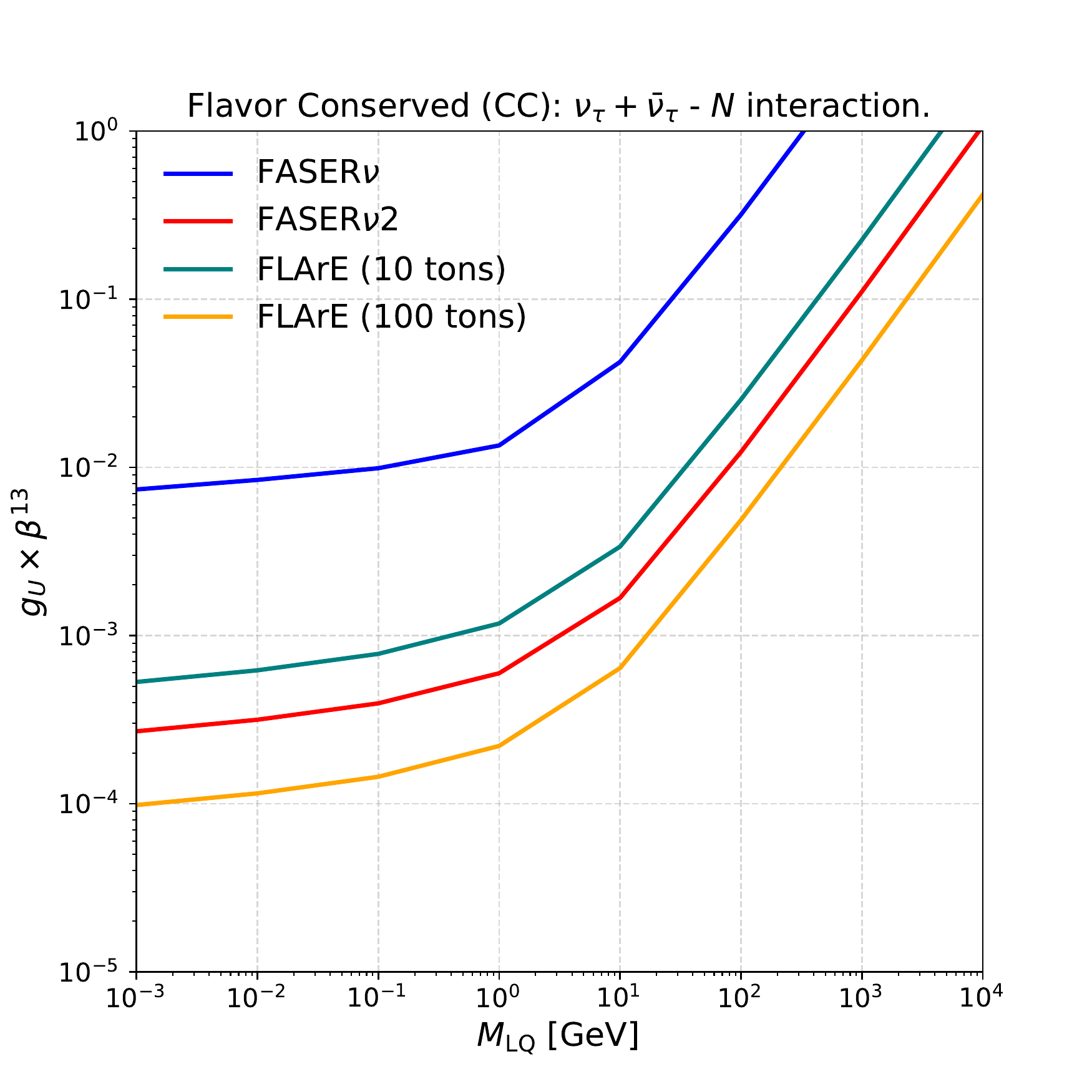}
\end{subfigure}
\hfill
\caption{95\% C.L. sensitivity curves on $g_{U}\times \beta^{1j}$, with $j=1$ (left), $j=2$ (middle) and  $j=3$ (right) for the CC scattering.
We do not include the lepton-flavor-changing processes in our 
calculations, i.e., only $\nu_{i}+N\to l_{i}+N'$ are considered.
}
\label{fig:CC_gUBeta}
\end{figure}

In the NC scattering, we cannot distinguish the neutrino flavor
and the leptoquark couplings $\beta_{L/R}^{1j}$ allow lepton-flavor
changing, we include all three flavors of neutrino
in the final state.
On the other hand, we can identify the outgoing
charged leptons in the final state in the CC scattering.
Since we know the flavor of the incoming neutrino fluxes, we 
consider only the lepton-flavor-conserving processes. Thus,
we can actually obtain the sensitivity $\beta^{1j}$ for each $j$ neutrino generation.

The sensitivities obtained using the CC scattering for each 
neutrino flavor are shown in \fig{fig:CC_gUBeta}.
Comparing the sensitivity of each coupling $g_U \times \beta^{1j}$, 
we can see the coupling for muon (second generation) has the best
sensitivity at the FPF detectors since the muon (anti)neutrino flux
is the largest, while the sensitivity for the tau (anti)neutrino 
is the least as the tau neutrino flux is the smallest. 
In \fig{fig:CC_gUBeta}, we can also see that FLArE (100 tons) has
the best sensitivity, followed by FASER$\nu 2$, FLArE (10 tons), and
FASER$\nu$.
Next, we include the lepton-flavor-changing processes
by including all leptons or antileptons in the CC final state. 
These lepton-flavor-changing contributions increase the number of
signal events. For simplicity we take all $\beta^{1j}_{L/R}$ equal to 
1. Thus, we obtain the sensitivities on $g_U$.

\begin{figure}[thb!]
\centering
\captionsetup{justification=centering}
\begin{subfigure}{0.48\textwidth}
\centering
\includegraphics[width=\textwidth]{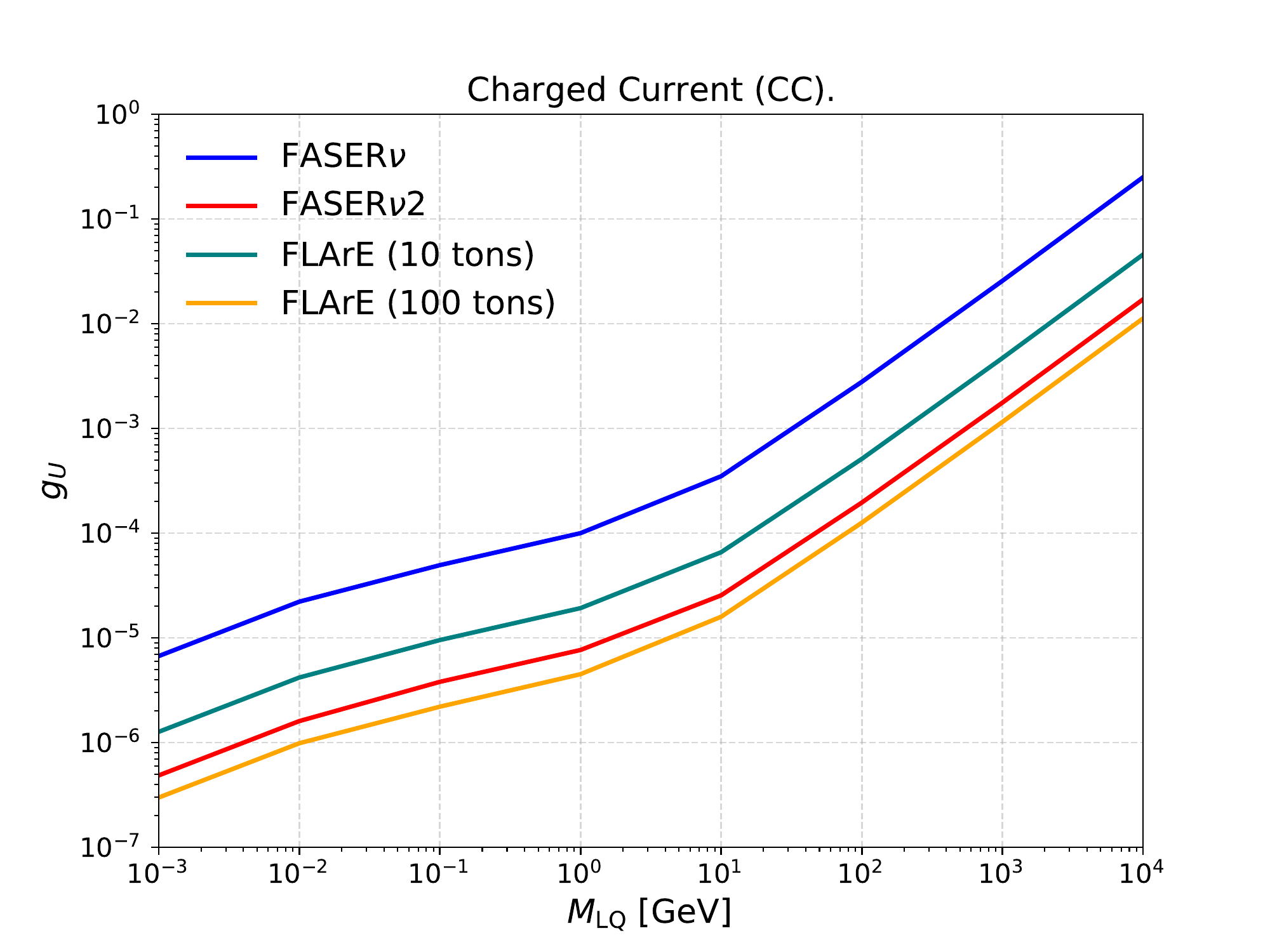}
\end{subfigure}
\hfill
\begin{subfigure}{0.48\textwidth}
\centering
\includegraphics[width=\textwidth]{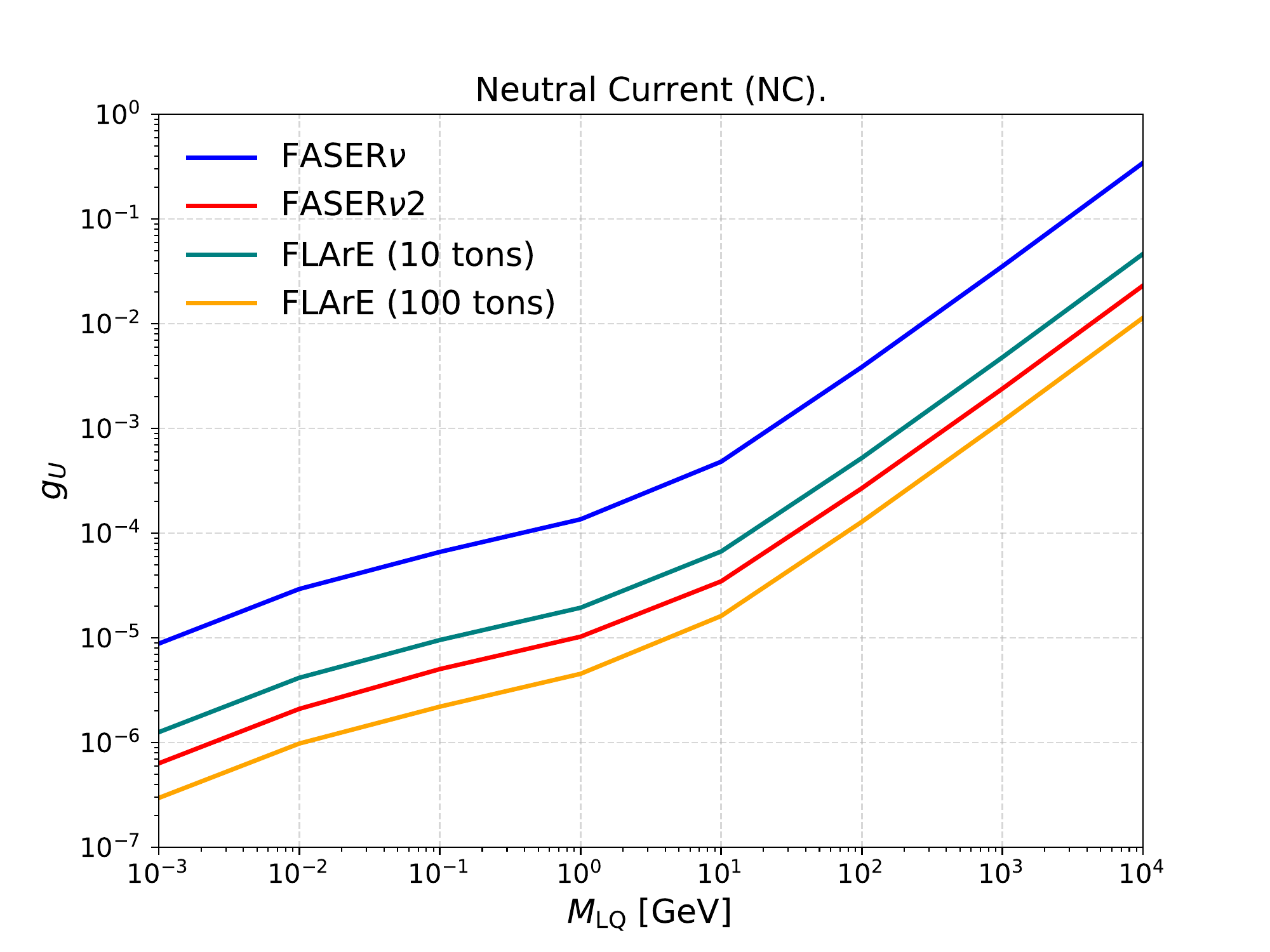}
\end{subfigure}
\hfill
\caption{95\% C.L. sensitivity curves on $g_{U}$ for 
CC: $\nu_{i}+N\to l_{j}+N'$ (left panel) and 
NC: $\nu_{i}+N\to \nu_{j}+N'$ (right panel).}
\label{fig:NCandCC}
\end{figure}

For NC scattering we include all (anti)neutrino flavors in the
final state, and thus we can obtain the sensitivities on $g_{U}$.
We show the results for FASER$\nu$, FASER$\nu$2, and FLArE 
in \fig{fig:NCandCC}. It is clear that FLArE (100 t) has the best
sensitivity, followed by FASER$\nu$2, FLArE (10 t), and FASER$\nu$.
Comparing the left (CC) and right (NC) results in \fig{fig:NCandCC},
the sensitivities obtained using CC scattering are slightly better 
than those using NC scattering. 
This is because we have summed over all charged leptons in CC and all
neutrino flavors in NC, and the CC suffers from the massive tau lepton
in the phase space. Thus, the sensitivities using CC are only
slightly better than those using NC.
The sensitivity curves are also marginally better than 
that of the $g_{U}\times \beta^{12}$ curves shown in the 
middle panel of \fig{fig:CC_gUBeta}, since the muon (anti)neutrino and 
flux dominates over that of electron and tau neutrinos in the FPF 
detectors.

\begin{figure}[t!h]
\centering
\captionsetup{justification=centering}
\includegraphics[width=\textwidth]{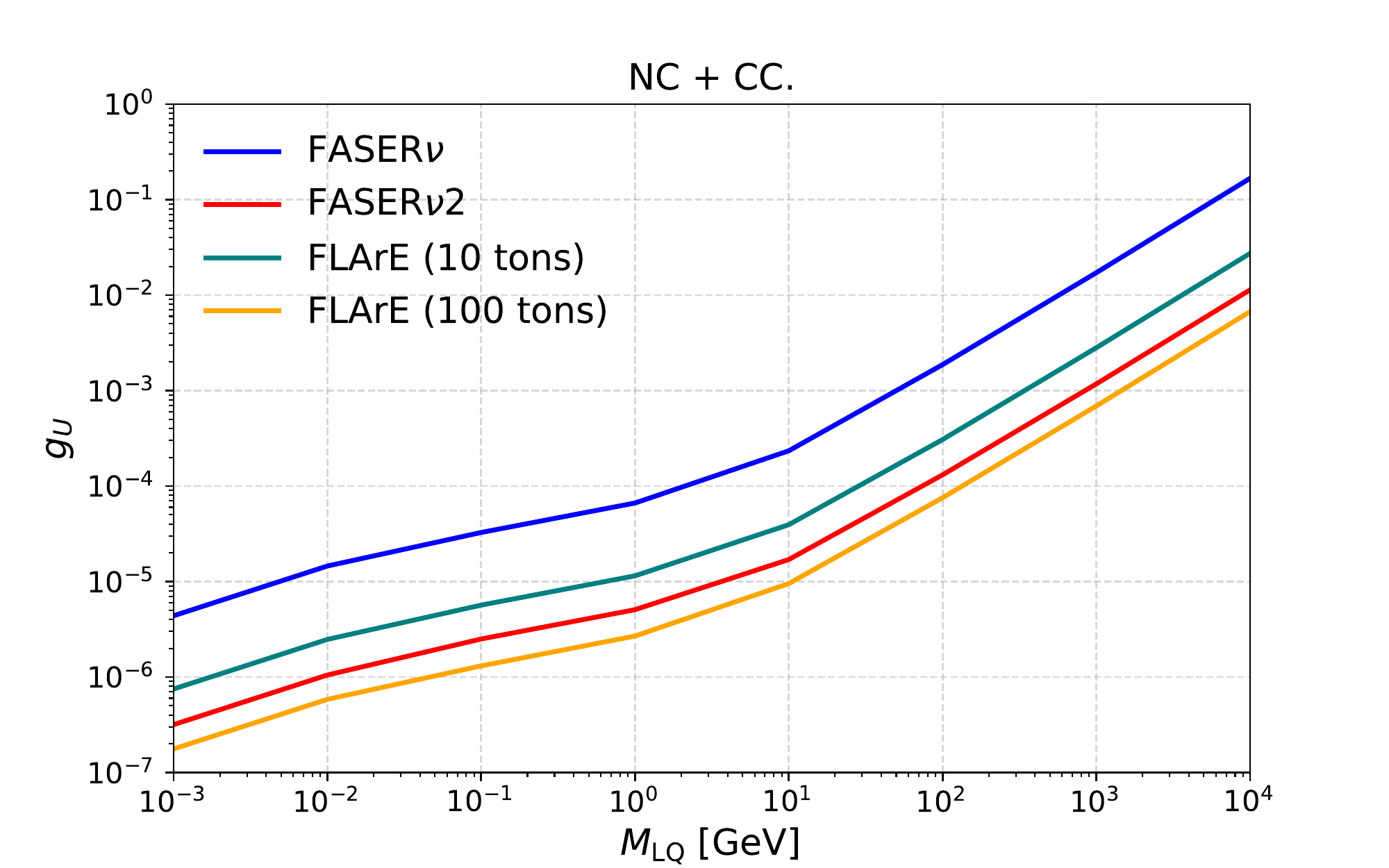}
\caption{95\% CL sensitivity curves on $g_{U}$ when combining both 
CC and NC events.}
\label{fig:gUAll}
\end{figure}

Finally, we can combine the number of signal events for both CC and NC
scattering. The sensitivity curves for the overall coupling $g_{U}$ 
are shown in \fig{fig:gUAll}. It clearly shows the improvement
from the individual NC or CC result.

The sensitivity can reach down to $4\times 10^{-5}$ for the current
FASER$\nu$ in the sub-GeV leptoquark mass range. In the electroweak 
leptoquark mass range, the sensitivity lies in between 
$10^{-3}$ and $10^{-1}$. For other proposed detectors, the sensitivity
can go down by 1 to 2 orders of magnitude. 
We can also compare the sensitivity of each FPF detector: 
FLArE 100 tons is the best, the next is the FASER$\nu$2 and then 
FLArE 10 tons. 

\begin{figure}[h]
\centering
\captionsetup{justification=centering}
\begin{subfigure}{0.47\textwidth}
\centering
\includegraphics[width=\textwidth]{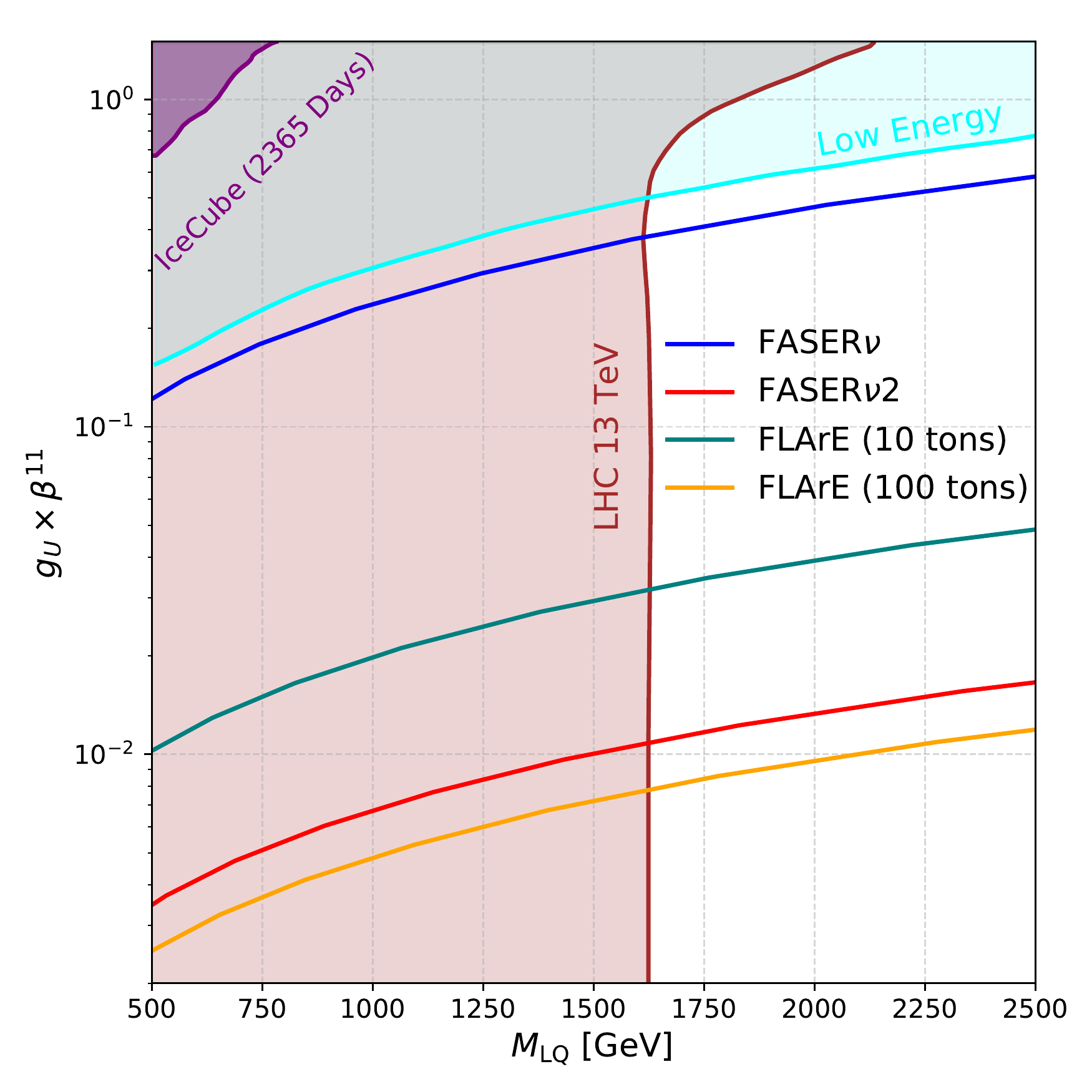}
\end{subfigure}
\hfill
\begin{subfigure}{0.47\textwidth}
\centering
\includegraphics[width=\textwidth]{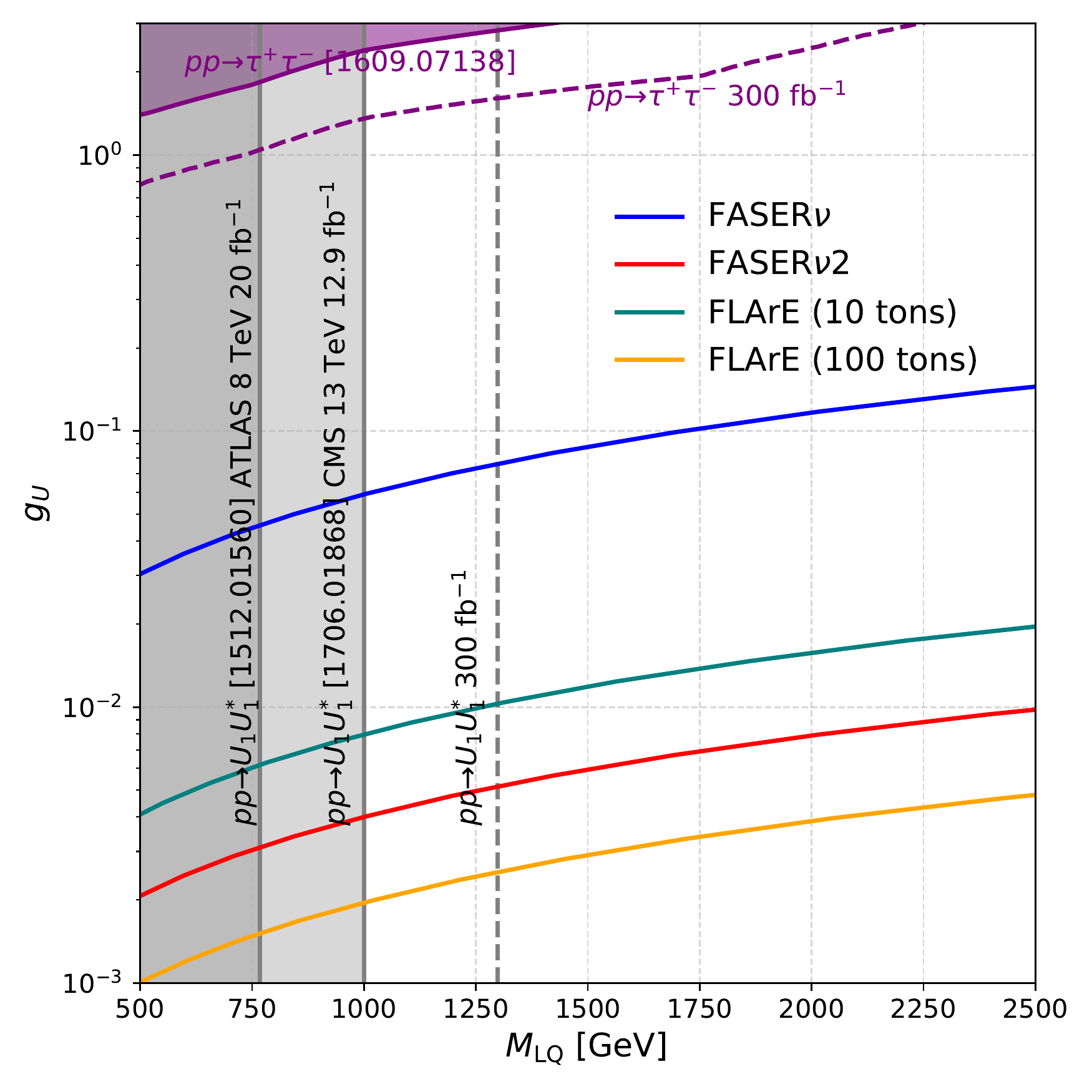}
\end{subfigure}
\hfill
\caption{Comparing the sensitivity reach from the FPF detectors with other 
constraints in the TeV regime. 
{\bf (Left)} Excluded regions for the $g_{U}\times \beta^{11}$ from IceCube 
\cite{IceCube:2013low, IceCube:2014stg, IceCube:2017zho,Schneider:2019ayi}, 
LHC 13 TeV \cite{ATLAS:2016wab, CMS:2015xzc}, and from low energy 
experiments \cite{Arnan:2019olv}. 
{\bf (Right)} Bounds on the overall coupling $g_{U}$ from 
ATLAS \cite{Faroughy:2016osc}, CMS \cite{DiLuzio:2017chi} for 
both LQ-pair and $\tau$-pair \cite{Faroughy:2016osc} production, 
and the projection for High Luminosity LHC.  }
\label{fig:Compare}
\end{figure}

In Figure \ref{fig:Compare}, we compare our results to those in the literature, 
especially in the LQ mass range from 500 GeV to 2.5 TeV (in linear scale
to match the mass region in \cite{Dorsner:2019vgp, Buttazzo:2017ixm}). 
In the left subfigure, the corresponding coupling ruled 
out by IceCube \cite{IceCube:2013low, IceCube:2014stg, IceCube:2017zho,Schneider:2019ayi} 
is above 0.6 - 1.0 for the first-generation quark-lepton coupling 
in the 0.5 - 1 TeV mass range, 
while the bound from low-energy experiments \cite{Arnan:2019olv} 
can go down to 0.2 - 0.8 for the 500-2500 GeV mass range.
The LHC-13TeV put a stringent limit for the LQ mass around 1600 GeV 
due to pair production, which is independent of the Yukawa couplings
as long as they are not extremely small as the current search is based on prompt decays of the 
leptoquarks \cite{Diaz:2017lit, Schmaltz:2018nls}.
However, for LQ above 1.6 TeV the first-generation coupling 
is only excluded in the region above 0.7.
In the right subfigure for the overall coupling, we show the
lower mass limits of the leptoquark from both 
ATLAS \cite{Faroughy:2016osc} and  CMS \cite{DiLuzio:2017chi}, 
and their projections for leptoquark-pair production, 
which exclude LQ masses below 1 TeV. We also included 
the bound from $\tau$-pair production from \cite{Faroughy:2016osc} 
and its projection. For higher LQ masses around 2 TeV, the limits for 
each individual coupling can be pushed down to around 0.1, 
even smaller than the results from recent study on SMEFT 
operators from the high-$p_{\mathrm{T}}$ tail 
in \cite{Allwicher:2022gkm}.

We overlay our results for FPF detectors in Fig.~\ref{fig:Compare}.
For FASER$\nu$, the sensitivity can reach down 
to $\sim$ 0.01 or $M_{U_{1}}=500$ GeV to $\sim$ 0.1 for 
$M_{U_{1}}=2500$ GeV. Other FPF detectors, as we showed before, 
can achieve better sensitivities than FASER$\nu$. 
Such FPF experiments can probe the currently allowed regions and 
improve the constraints on the vector leptoquark model.

The sensitivity curves from FASER$\nu$ are already better 
than those excluded regions by $1-2$ orders of magnitude. 
Furthermore, the other FPF detectors can further improve the
sensitivity by another $1-2$ orders of magnitude.
in the TeV leptoquark mass range. 
In addition, the experiments proposed in FPF can probe a broad 
mass range of the leptoquark; especially, the FPF detectors 
are able to probe the region with small couplings in the sub-GeV 
leptoquark mass range, which is a challenge for the 
conventional hadron colliders.

\section{Conclusions}

The proposed Forward Physics Facility offers an array of 
experiments, which can take advantage of the unique neutrino beam in
the energy range of a few hundred GeV to TeV to explore the physics 
beyond the SM. We have investigated the sensitivity reach on the
leptoquark model at a number of experiments, including FASER$\nu$, 
FASER$\nu2$, FLArE(10 tons), and FLArE(100 tons). We compared
the advantage of the FPF experiments in a wide mass range of
the LQ mass and to determine the flavor dependence of the 
couplings between the neutrinos and this LQ.

We have covered a wide mass range of LQ mass $10^{-3}~\rm GeV\leq M_{LQ}\leq10^4~GeV$ in our study. Among all the proposed FPF
experiments, FLArE(100 tons) has the best sensitivity to the LQ model,
whereas FASER$\nu$ has the least. The sensitivity curves for all the experiments follow a similar pattern, in which the sensitivity is
weakened with the increment of the LQ mass. The unique feature of 
the LQ is that it contributes to both NC and CC-like 
neutrino-nucleon scattering at the FPF.  We obtained
the final sensitivities for the LQ couplings by combining both the 
CC and NC events.

\section*{Acknowledgement}
Special thanks to Felix Kling and Zeren Simon Wang for enlightening discussions. T.T.Q.N. would like to thank the Department of Physics and Center for Theory and Computation, NTHU, Taiwan for its hospitality.
The work of T.T.Q.N. as a research assistant is supported in part by the Ministry of Science and Technology (MoST) of Taiwan under Grant No 111-2112-M-001-035. K.C. and C.J.O. are supported by MoST under 
Grant no. 110-2112-M-007-017-MY3.

\bibliographystyle{JHEP.bst}
\bibliography{biblio.bib}
\end{document}